\documentstyle[12pt,epsf]{article}
\newcommand{\beq}{\begin{equation}}
\newcommand{\beqa}{\begin{eqnarray}}
\newcommand{\eeq}{\end{equation}}
\newcommand{\eeqa}{\end{eqnarray}}
\newcommand{\MM}{\bf \cal M}
\newcommand{\g}{\gamma}
\newcommand{\G}{\Gamma}

\newcommand{\sg}{\sigma}

\newcommand{\pr}{\partial}

\newcommand{\del}{\delta}

\renewcommand{\L}{\Lambda}
\renewcommand{\b}{\beta}
\renewcommand{\a}{\alpha}
\newcommand{\n}{\nu}
\newcommand{\m}{\mu}

\newcommand{\eps}{\;\epsilon\;}
\newcommand{\ep}{\varepsilon}

\newcommand{\r}{\prime}
\newcommand{\DD}{\cal D}

\begin{document}

\topmargin 0pt
\oddsidemargin 5mm
\headheight 0pt

\topskip 5mm

%\addtolength{\baselineskip}{0.5\baselineskip}

\begin{flushright}
NBI-HE-96-35\\
July 1996\\
Revised January 1997
\hfill
\end{flushright}

\begin{center}
\hspace{10cm}

\vspace{15pt}
{\large \bf
THE
METRIC ON THE SPACE
OF YANG-MILLS CONFIGURATIONS }

\end{center}

\vspace{20pt}
 
\begin{center}
{\bf Peter Orland}\footnote{Work supported 
by PSC-CUNY Research Award Program 
grants
nos. 666422 and 667438.}

\vspace{15pt}
 
The Niels Bohr Institute \\
Blegdamsvej 17, DK-2100 \\
Copenhagen {\O}, Denmark \\
orland@alf.nbi.dk\\
\vspace{8pt}

and\\

\vspace{8pt}

The City University of New York\\
Baruch College\\ 
and\\
The Graduate School and University Center\\
New York, NY \\
orland@gursey.baruch.cuny.edu
\footnote{Permanent address}\\

\end{center}
 
\vspace{10pt}

%\addtolength{\baselineskip}{0.5\baselineskip}

\begin{center}
{\bf Abstract}

\end{center}

A distance function on the set
of physical equivalence classes of
Yang-Mills configurations considered by Feynman and
by Atiyah, Hitchin and Singer is studied
for both the
$2+1$ and $3+1$-dimensional 
Hamiltonians. This 
set equipped with this distance
function is a 
metric space, and in fact
a Riemannian manifold as 
Singer observed. Furthermore, this
manifold is complete. Gauge configurations
can be used to parametrize the 
manifold. The metric tensor without
gauge fixing has zero eigenvalues, but 
is free of ambiguities on the entire
manifold. In $2+1$ dimensions the problem
of finding the distance
from any configuration to a pure gauge configuration
is an integrable system of two-dimensional
differential equations.

A calculus of
manifolds with singular
metric tensors is developed and the Riemann curvature is calculated
using this calculus.  The Laplacian on Yang-Mills wave functionals
has a slightly different form from that claimed earlier.

In $3+1$-dimensions there are field configurations
an arbitrarily large distance from a pure gauge configuration with
arbitrarily
small potential energy. These configurations
resemble long-wavelength gluons. Reasons why there nevertheless
can be a mass gap in the quantum theory are proposed.

\vfill
\newpage

\section{Introduction}

The Schr{\"o}dinger representation is certainly the most
powerful for visualizing and solving
finite-dimensional quantum
systems. This is undoubtedly why it is emphasized more
than the Heisenberg and interaction
representations, even in
advanced textbooks on quantum mechanics. It is
is just as useful in field theory 
as the interaction representation or
path integrals for developing
the rules of perturbation theory (though somehow
this fact 
is not as often
emphasized in textbooks) \cite{rmp}. There is also a 
conceptual value to the notion of a quantum wave 
functional \cite{schrod}. For a long
time there has been the hope that a variational
method might give a reasonable approximation to the vacuum
of the Yang-Mills Hamiltonian \cite{vari}. Even should
this turn out not to be 
the case, wave functionals of field configurations are easier
to visualize than the abstact states of the Heisenberg 
representation and are worth investigating further.

Serious 
attempts to understand the long-distance physics of Yang-Mills
theories in the
Schr{\"o}dinger representation
were made by Feynman (in $2+1$ dimensions) \cite{feynman}
and by Singer \cite{singer}. These papers concerned 
the geometrical properties
of the space of the physical
configurations of the gauge field. This 
infinite-dimensional space is different in character from that
of simpler
field theories. In particular, two gauge
fields which differ by a gauge transformation (one
has the {\em option} of insisting that this gauge
transformation is of Chern-Simons
number zero in $3+1$ dimensions) must be identified
as the same point of configuration space. 

Feynman discussed
the notion of distance between two field
configurations \cite{feynman}. He attempted to
minimize this distance by making
gauge transformations. In this way, he obtained an estimate
for the distance between a pure gauge configuration and a configuration
with a constant magnetic field. The point
of doing this was to estimate the electric energy
of certain wave functionals and thereby to minimize the total
energy. On the basis of this result, he gave a conjecture
for the lightest glueball state of the $2+1$-dimensional theory. In
fact, the minimal distance had been already discussed by Atiyah, Hitchin
and Singer, who used it to calculate the dimension
of the moduli space of instantons \cite{atiyah}.

Singer defined an inner product
on the configuration
space and discussed how to regularize
tensor contractions so that the eigenvalues
of the Ricci tensor and
the Laplacian (which is the kinetic term of the Yang-Mills
Hamiltonian) were well-defined \cite{singer}. The first work
along similar lines
can be found in the
papers by Babelon, Viallet, Mitter, Narasimhan and Ramadas
\cite{babelon1, babelon2}. In particular, Babelon
and Viallet studied geodesics on this space
(and thereby classical dynamics
with no potential energy) \cite{babelon2}. A difficulty 
encountered
in \cite{babelon2} was that geodesic motion appeared 
ill-defined at points of the manifold where the
operator ${\cal D}^{2}$ has zero eigenvalues.

Recently an important advance was made by 
Karabali and Nair \cite{karnair1}, who discovered
a gauge-invariant description of Yang-Mills theories in 
$2+1$ dimensions. Their method solves the constraint of Gauss' law and
they have been able to show the existence of a gap at 
strong coupling. They have also made some arguments for the
persistence of this gap at any coupling \cite{karnair2}.

In this article, the configuration space of the
Yang-Mills 
theory is investigated using general
methods, some
of
which are not specific to the number of dimensions. The issue
of mass generation is confronted, but not that of
color confinement.

After introducing the notation and reviewing the relevant
physics in section 2, a heuristic way to
eliminate Gauss' law from the $A_{0}=0$ Yang-Mills theory
is discussed in section 3. The methods used to study
gauge theories in references
\cite{singer, babelon1, babelon2} start 
from the same standpoint. The 
difficulty raised in \cite{babelon2}
concerning zero modes of the covariant
operator ${\cal D}^{2}$ is shown. Because of this
problem, such a formalism cannot be accepted
on faith, and it is carefully
justified in
sections 4, 5 and 6. In the correct metric
tensor, these zero modes do not present a problem.

A distance function
motivated by Feynman's analysis \cite{feynman}
is investigated and found to
define
a metric space in section 4. This
distance function coincides with that of
Atiyah et. al. \cite{atiyah}, though it will not
be considered only on self-dual connections. Furthermore, with
this metric, the space of physical
configurations is complete, meaning roughly that there
are no ``holes" in this space. In section 5, it 
is found that this distance
function is also the ``best" choice, in that the distance
between two points is the length of a minimal curve connecting
these points. In fact, this property
had been conjectured by Babelon and Viallet \cite{babelon2}. In 
section 6, the 
distance function is used to define
the Riemannian manifold where the gauge connection itself
is used as the coordinate. This is a very
different choice
of coordinates from that taken
in references \cite{karnair1, karnair2, freedman}. It 
is shown that this can be done
everywhere on the manifold, in
spite of the
fact that the metric tensor has zero
eigenvalues. The problem
with zero modes of ${\cal D}^{2}$ \cite{babelon2}
is completely
resolved. Since singular metric 
tensors are rarely used, a review
and a 
simple finite-dimensional
example are given in the appendix. Gauss' law 
can be solved in any dimension
of space-time. The
Hamiltonian 
is not local (this is also
the case for a physical gauge condition, such as the
axial gauge, and also in a polar representation of the 
gauge field \cite{simonov}). For non-infinitesmal distances, it
is hard to calculate the distance function explicitly, but
in section 7 it is shown 
that for some interesting
cases in $2+1$ dimensions its calculation can be reduced to the
solution
of an integrable set of differential equations.

There are various theorems relating 
the eigenvalues of Ricci tensor to the spectrum of the Laplacian 
\cite{davies}. Ultimately, these
results may be of value to the Yang-Mills Hamiltonian. The
validity of these theorems usually 
depends upon the manifold having compact closure. This is 
probably not true of the unregularized
Yang-Mills configuration space, but is 
certainly so
if the theory is put
on a finite lattice \cite{kudinov}. In section 8 
the curvature tensor
is found at points
where zero
modes are not present \cite{singer, babelon2}, but 
no application of these
results is made to the spectrum.

Once the metric on the 
manifold of Yang-Mills configurations is known, the
kinetic-energy operator (in other
words the electric energy) can be obtained. This 
operator, the coordinate-invariant Laplacian, is found
in section 9. The result
differs in two respects from that obtained by previous
authors \cite{singer, babelon2}. First
of all, the zero
modes of ${\cal D}^{2}$ present
a complication, which leads to a 
slightly different metric tensor. Secondly, even
with the correct
metric tensor, an additional
term is present in the Laplacian, 
which had been missed in references
\cite{singer, babelon2}.

In section 10 some
unusual
aspects of the magnetic energy 
functional are investigated for different
dimensions of space-time. In
$3+1$ dimensions, for at least some regions of
the manifold of configurations, 
the ``hills and valleys" of the potential energy
are extremely steep. In particular, there are regions
a small geodesic distance from any pure gauge of extremely
large potential energy. What is even
more striking is that there exist points on the manifold
of arbitrarily large geodesic distance from any pure gauge, yet 
possess arbitrarily small potential
energy; this is very surprizing
for a theory which is expected
to have a mass gap in the spectrum. In contrast, the 
potential energy
of the
$2+1$-dimensional model can have {\em any} positive
value
for configurations a fixed distance from the
pure gauge configuration. Discussion of the origin
of the mass gap for $2+1$ and $3+1$ dimensions (in
spite of the difficulty noted above!) is in
Section 11.

The philosophy here differs significantly
from that of most
people currently working in this field
\cite{zwanziger, van baal, cutkosky, fujikawa, fuchs}
in one significant             
respect. The space of Yang-Mills connections (discussed
here for only connections in space with $A_{0}=0$, rather
than for the Euclidean path integral) is not {\em restricted}
to configuration space
(or orbit space) by some sort of gauge fixing. Instead the points
of configurations space are defined as equivalence 
classes of connections, and no restrictions whatever are imposed
on these connections. In this sense, the approach
here is similar in spirit to that of 
Friedberg et. al. \cite{friedberg}. The problem of
finding the fundamental region (or the interior
of the Gribov horizon) has been traded for
a singular metric tensor. Metric tensors with
vanishing eigenvalues introduce subtle
new features into 
differential geometry, but as shown in the
appendix, they present no
inconsistencies. So-called
reducible connections \cite{fuchs}, i.e. gauge
configurations which are invariant under
a subgroup of gauge transformations present
no difficulty in defining the Laplacian
on configuration space. It is argued
that they are also not 
relevant in the
calculation of inner products between
certain wave functionals (though
this issue needs to be
examined more carefully). This
will be briefly discussed at the end of section 6.

Since the 
literature on the configuration space of Yang-Mills 
gauge theories
is by now quite extensive, it is important to say at the
outset what in this paper can or cannot be found elsewhere. The
ideas discussed in section 3 are
not new and can be found in references
\cite{singer, babelon1, babelon2}. While 
many of the ideas of sections 4, 5 and 6 can be found in 
references 
\cite{feynman, singer, atiyah, babelon1, babelon2}, the 
ideas are tied
together in a way which is entirely new and
the main results are slightly different. Most importantly, it 
is shown that the metric tensor and geodesic trajectories
are well-defined {\em even at
so-called
non-generic points}
(where there are zero modes of ${\cal D}^{2}$). Singer presented
the curvature tensor for
other points in his paper \cite{singer}, as
did Babelon and Viallet \cite{babelon2}. This tensor
is calculated using the 
methods introduced in this paper in section 8, though
the result of this calculation is not new. The kinetic
term of the Yang-Mills theory obtained in section 9
differs in certain respects from that claimed earlier 
\cite{singer, babelon2}. As far as I
know, none of the results
presented in sections 7, 9, 10 and 11 can be found
elsewhere.

As
I have
found the properties of the manifold of physical Yang-Mills configurations
hard to visualize, I have tried to use careful definitions and theorems
in sections 4, 5 and most of section6, so 
as not to prove wrong results. These sections
use some basic concepts of real analysis and integration theory, but are
hopefully accesible to those
physicists who are not well-versed in these concepts. No
attempt to be completely
rigorous is made in
rest of the paper.

\section{Preliminaries}

The main purpose of this section is to establish
some notation, conventions and definitions which will be used
in the remainder of the article. Those initiated into the
Hamiltonian formulation
of gauge theories will not find anything particularly
new until section 3.

The dimension of space-time is denoted by $D+1$. Let 
$t_{a}$ be an orthonormal basis for the fundamental repesentation
$su(N)$, the 
Lie algebra of the group $SU(N)$. This basis
can be chosen so that $tr\,t_{a} t_{b}=\delta_{a\,b}$
and $[t_{a}, t_{b}]={f}^{a\,b\,c} \,t_{c}$. A gauge field $A$ is 
an
$su(N)^{D}$-valued vector field $A_{j}(x)$, $j=1,...,D$ 
on a $D$-dimensional
manifold $M$, whose points
will be denoted by letters $x, y, z$. The covariant
derivative operator for the gauge field $A$ is 
$D^{A}_{j}=\pr_{j}-iA_{j}$, which will sometimes just
denoted by $D_{j}$, depending on the context. The adjoint
representation
of the covariant derivative
is 
${\cal D}_{j} \equiv [D_{j}, \cdot]$. The curvature or 
field strength is $F^{A}_{jk}=-i[D^{A}_{j}, D^{A}_{k}]$. 

A gauge transformation is any differentiable mapping 
of $M$ into the fundamental representation of 
$SU(N)$. The space of gauge transformations
will be written $\em G$ (in other words $h \eps {\em G}$ means
$h(x)\in SU(N)$). Let 
$A_{\m}(x)=A_{\m}^{a}(x) \,t_{a} \in su(N)$ be a gauge 
field. If $B_{\m}(x)=B_{\m}^{a}(x) \,t_{a} \in su(N)$
such that $A$ can be changed to $B$ by a gauge transformation 
$h \in {\em G}$, i.e.
$h(x)^{-1}\,D^{A}_{\m}(x)\,h(x)=D^{B}$, then $A$ and $B$ will be said
to be gauge equivalent, sometimes expressed
as $B=A^{h}$ (a more precise definition of gauge equivalence will be given 
in section 4). Clearly gauge equivalence is an 
equivalence relation. For $D\neq3$ two
gauge-equivalent fields will also said to be physically eqivalent.

For the case of $D=3$ a subtlety arises which permits (but
does not require)
the definition
of a different equivalence 
relation \cite{thetavac}. With this
definition, a pair of gauge-equivalent gauge 
fields are not neccesarily
physically 
equivalent. Suppose that $M$ is taken to be
$R^{3}$ with all gauge fields falling off at infinity
as 
\beqa
A_{j}(x) \longrightarrow -i\;h(x)^{-1} \pr_{j} h(x)+O(1/x) \;. \nonumber
\eeqa
A gauge transformation $h\eps {\em G}$ has 
an associated Chern-Simons 
number
\beq
\n[h]=\frac{i}{12\pi^{2}} \int d^{3} x \; \ep^{i\;j\;k} \;
tr\;h(x)^{-1}\pr_{i} h(x)
\;h(x)^{-1}\pr_{j} h(x)\;
h(x)^{-1}\pr_{k} h(x) \;, \label{2.01}
\eeq
which is an integer. One has
the option of insisting
that two gauge configurations $A$
and $B$
are physically equivalent only if 
$h(x)^{-1}\,D^{A}_{\m}(x)\,h(x)=D^{B}$, with $\n[h]=0$. Now
if $h, f \eps {\em G}$ are two gauge transformations, then
$\n[hf]=\n[h]+\n[f]$. This property means that :\begin{itemize}
\item $\n[h]=\n[f]$ is
an equivalence relation and therefore the
gauge transformations
with a particular Chern-Simons number $\n$ constitute an equivalence
class ${\em G_{\n}}$ and
\item physical equivalence of gauge fields is an equivalence relation. 
\end{itemize}
Gauge configurations have the Chern-Simons integral
\beqa
C[A]=\frac{1}{12\pi^{2}} \int d^{3} x \; \ep^{i\;j\;k}
tr(A_{i}F_{j\;k}+\frac{2i}{3}A_{i}A_{j}A_{k}) \;, \label{2.02}
\eeqa
which is changed under a gauge transformation satisfying (\ref{2.01}) by
$C[A] \rightarrow C[A^{h}]=C[A]+\n[h]$. Thus the Chern-Simons integral
is the same for each element of a physical-equivalence class making it
sensible to write $C[A]=C[\a]$. 

The Hamiltonian
of the $SU(N)$ Yang-Mills theory in $A_{0}=0$ (temporal)
gauge in 
the Schr\"{o}dinger picture is
\beq
H= \int_{M}\,  d^{D}x\; 
[-\frac{e^{2}}{2} \,tr\,\,\frac{\delta^{2}}{\delta\, A_{j}(x)^{2}}
+ \frac{1}{4e^{2}}  \,tr\,F_{j\,k}(x)^{2}]\;. \label{1.1}
\eeq  
The allowed 
wave functionals $\Psi$ satisfy the condition that if $A$ and $B$
are physically equivalent
\beq
\Psi[A]=\Psi[B]\;. \label{1.2}
\eeq

The 
space of gauge field configurations (or connections)
will be denoted
by ${\cal U}$. The physical configuration space (sometimes
called the orbit space) 
is then ${\cal U} / G \equiv {\MM}_{D}$ when $D\neq3$ and ${\cal U} / G_{0}
\equiv {\MM}_{3}$ when $D=3$. In other
words, for $D\neq 3$, configuration space, ${\bf \cal M}_{D}$ is 
defined to be the set of 
gauge-equivalence
classes of gauge fields (these were already defined
to be physically equivalent). For $D=3$, configuration 
space ${\bf \cal M}_{3}$ is defined to be the set of 
physical-equivalence
classes of gauge fields. The elements
of ${\bf \cal M}_{D}$ will be called physical configurations. The 
physical configurations containing the
gauge fields $A, B, S,...\in {\cal U}$ will be respectively
denoted $\a, \b, \sg,...\in {\bf \cal M}_{D}$. Thus, $A$ is
physically
equivalent to $B$ if and only if 
$\a=\b$. By (\ref{1.2}), wave functionals depend
only on physical configurations ($\a$ in ${\MM}_{D}$), not
gauge configurations ($A$ in ${\cal U}$).

\section{Projecting out gauge transformations}

Can one modify the form of the Hamiltonian (\ref{1.1}) so that
imposing (\ref{1.2}) is unneccesary? A simple guess is to change
the kinetic term to obtain
\beqa
H= -\int_{M}\,  d^{D}x \;
\frac{e^{2}}{2}\,tr\,\frac{\delta}{\delta\, A_{j}(x)}\;
G^{-1}_{j\,k} \; \frac{\delta}{\delta\, A_{k}(x)} \nonumber
\eeqa
\beq
+ \int_{M}\,  d^{D}x\; \frac{1}{4e^{2}}  \,tr\,F_{j\,k}^{2}(x) \;, \label{1.3}
\eeq
where
\beq
G^{-1}_{j\,k}=\delta_{j\;k} 
-{\cal D}_{j} \frac{1}{{\cal D}^{2}} {\cal D}_{k}
\;, \label{1.33}
\eeq 
is a projection operator (being very sloppy about domains of covariant
derivative operators) which 
vanishes on infinitesmal 
gauge transformations $h(x)\approx 1-i\omega$, that is $G^{-1}_{j\,k} 
{\cal D}_{k}\omega=0$. For $D=3$, $G^{-1}$ neccesarily vanishes on 
gauge transformations whose Chern-Simons number is zero.

The reader may have noticed
a serious problem. The operator ${\cal D}^{2}$
can have eigenvectors with zero eigenvalues \cite{babelon2}. How
should the Green's function $1/{{\cal D}^{2}}$ in
(\ref{1.33}) be interpreted in this case?

It
is useful to introduce a generalization of the inverse of an operator. If
$\Theta$ is any operator on a space with orthonormal basis $\{|X)\}$ 
and $\Theta |X)=\lambda_{X} |X)$, then the operator and its generalized
inverse have the spectral representations (again being very careless about 
domains)
\beqa
\Theta=\sum_{X}\lambda_{X} |X)(X|\;, \nonumber \\
\Theta^{-1}=\sum_{X,\lambda_{X} \neq 0 } \lambda_{X}^{-1} |X)(X|\;. \label{1.5}
\eeqa

In the event
that ${\cal D}^{2}$ has zero modes \cite{babelon2}, its inverse
can be made well-defined using (\ref{1.5}). However, it
is not obvious that this is the correct prescription. It will
be shown in section 6 that precisely (\ref{1.5}) must be
used to define the inverse of ${\cal D}^{2}$. This means that (\ref{1.33})
is incorrect and must be replaced by
\beq
G^{-1}_{j\,k}=\delta_{j\;k} 
-{\cal D}_{j}\; {\cal P} \frac{1}{{\cal D}^{2}}\; {\cal D}_{k}
\;, \label{1.4}
\eeq 
where ${\cal P}$ denotes the principal value. This
is equivalent to using
(\ref{1.5}) to define the inverse
of ${\cal D}^{2}$. This may also be written as
\beqa
{\cal P} \frac{1}{{\cal D}^{2}} = \frac{1}{2} 
(\frac{1}{{\cal D}^{2}+i\epsilon}
+\frac{1}{{\cal D}^{2}-i\epsilon}) \;. \nonumber
\eeqa

The operator $G^{-1}$ defined
in (\ref{1.33}) or (\ref{1.4})
is formally idempotent (it is
equal to its square), hence, if it is 
meaningful, each of its 
eigenvalues can only be either
zero or one. It does not have an inverse except in
the sense of (\ref{1.5}). According to this definition $G^{-1}$ is its own 
inverse, $G=G^{-1}$; it plays the
role of the identity on the space of connections modulo small
gauge transformations.

A conjecture for
the
kinetic
term is then:
\beqa
T=-\frac{e^{2}}{2}\,
(\frac{\delta}{\delta\, A})^{T} G ^{-1}
\frac{\delta}{\delta\, A} \;, \label{1.55}
\eeqa
This resembles a covariant 
Laplacian on a Riemannian
metric space 
with metric tensor $G$, since the 
determinant of the idempotent operator G, upon
removal of zero modes, is equal 
to one and $G=G^{-1}$. It turns
out, however, that (\ref{1.55})
is still the wrong answer, even with the
correct metric (\ref{1.4}). There is another
term present.

Before taking
these ideas seriously, several problems need
to be dealt with. The problem of
inverting ${\cal D}^{2}$ \cite{babelon2} must
be solved. It needs to be verified
that 
the complicated operator (\ref{1.4}) 
has no
zero modes other than small
gauge transformations; other differential
operators certainly do \cite{gribov}. These problems
will be solved in section 6. The Laplacian
needs to be determined, as will be done in section 9. It is 
perhaps noteworthy
that
the Hamiltonian of the form (\ref{1.3})
is easily shown to be correct for an Abelian gauge theory.

The reader may wonder 
why is is really valuable to study the
Laplacian in the approach
of this paper and references 
\cite{feynman, singer, babelon1, babelon2}, when 
a perfectly good 
definition of the quantized
Yang-Mills theory already exists; the lattice gauge theory. On
the lattice, at strong coupling, it 
is possible to implement Gauss' law to 
obtain only physical states, at least on a case
by case basis; they are strings. However, even
with a lattice theory, it is difficult to
get much intuition about the weak coupling
spectrum. An axial gauge eliminates the Gauss' law
constraint, but at the cost of introducing difficult
boundary conditions and an unmanageable kinetic
term. The methods
introduced here are not intended to compete
with the lattice, but rather to have a better understanding
of the geometry of the manifold ${\MM}_{D}$ and the potential
energy function on this manifold. A lattice discussion should
be very useful and will appear in reference \cite{kudinov}.

In
the next section a distance function will be defined; in section 6
it will be shown that the infinitesmal form of this
distance function exists and that the metric tensor is $G$, as defined
in (\ref{1.4}). There are various theorems relating 
the eigenvalues of Ricci tensor to the spectrum of the Laplacian 
\cite{davies}. Ultimately, these
results may be of value to the Yang-Mills Hamiltonian. The
validity of these theorems usually 
depends upon the manifold having compact closure. This is 
probably not true of the unregularized
Yang-Mills configuration space, but is 
certainly so
if the theory is put
on a finite lattice \cite{kudinov}. The curvature is
is found at points
where zero
modes are not present \cite{singer, babelon2} 
in section 8. The Laplacian is determined in section 9.

\section{The metric space of configurations}

This section and sections 5 and 6
presuppose some knowledge
of unsophisticated
aspects of 
analysis on the part of the reader, such
as completeness, total boundedness, compactness, measure and Hilbert 
spaces \cite{reed}. However, it is hoped that
most theoretical
physicists, who usually do not feel the need of 
such concepts (and who perhaps have happily
forgotten
them) will be able to follow the line of
argument. The rigorous part of this article begins here (it ends in
section 6).

First a distance
function will be defined \cite{feynman, atiyah}. Feynman 
found an estimate
of this distance function between the pure gauge configuration
and a configuration of constant magnetic
curvature for $D=2$. The
discussion here will be
more general, and it will be shown that ${\MM}_D$ equipped
with this function is a metric space. This
fact has already been noted for the case of self-dual
connections \cite{atiyah}. The value
of establishing this is that the distance
function must be therefore
a continuous function of the physical
configurations. This is one of the essential
ingredients needed to 
show that the manifold can be equipped with
a Riemannian metric. Furthermore, this
metric space is complete. It will 
be shown
in section 6 that the metric tensor defined with this
distance function is (\ref{1.4}). A general reference on 
metric spaces from the geometric point of view is
the book by Aleksandrov and Zalgaller \cite{alex}.

Some more careful definitions are needed. The
space of connections $\cal U$ is  redefined
to contain only those gauge fields which are Lebesgue
measurable, and are square-integrable, i.e.
\beqa
\int_{M} d^{D}x 
\;\sum_{k=1}^{D}\; tr\;A_{k}(x)^{2} < \infty\;, \nonumber
\eeqa

Physically, one would like to also restrict the connections
to those
whose field strengths exist and are square-integrable, i.e.
\beqa
\int_{M} d^{D}x 
\;\sum_{j,k=1}^{D}\; tr\;F_{j\;k}(x)^{2} < \infty\;. \nonumber
\eeqa
However, making this further restriction will ruin the completeness
property of the space of 
connections and is quite incovenient, as discussed
in the next paragraph.

Two
gauge fields $A$ and $B\in {\cal U}$ are identified 
if they are the same except on a set of measure zero. In other
words, no distinction is made between gauge fields which
are the
same almost everywhere. Thus the space
${\cal U}$ is a Hilbert space, and gauge fields
are representative of vectors in this Hilbert space. Furthermore, $G$ and 
$G_{\n}$ do not consist
of any $SU(N)$ valued functions $g(x)$, but only
those which
are differentiable and for which 
$ig^{-1} \partial g \in {\cal U}$. This
means in particular that
\beqa
\int_{M} d^{D}x 
\;\sum_{k=1}^{D}\; tr\;[ig(x)^{-1}\partial_{k} g(x)]^{2} < \infty. \label{gt}
\eeqa
Any element of ${\cal U}$ is mapped into another
element of ${\cal U}$ by such a gauge transformation. Notice 
that two 
almost-everywhere-equal gauge 
fields are still equal almost everywhere
after such a gauge transformation is appiled to both. Restricting the 
integral of the magnetic field strength squared destroys the completeness
property, and the space of connections is no
longer a Hilbert space. However, the
space of such restricted connections
is dense in ${\cal U}$. Therefore it not
neccesary to make this restriction. The situation is
similar to that one encounters when studying unbounded 
operators (the subspace of a Hilbert space which is acted
upon by 
an
unbounded operator is dense in the 
Hilbert space \cite{reed}).

It is not quite enough to say that two vectors in $\cal U$ are 
gauge-equivalent
if one can be gauge transformed to another by some gauge transformation
satisfying (\ref{gt}). The equivalence classes must actually be made
larger in order to obtain a metric space. Instead two vectors in $\cal 
U$ with representatives $A$ and $B$
will be said to be gauge-equivalent if there is a sequence of 
gauge transformations $g_{1}$, $g_{2}$,... each satisfying (\ref{gt}), such
that
\beqa
B=\lim A^{g_{n}} \label{equiv}\;,
\eeqa
in the usual metric of the Hilbert space {\cal U}, i.e. the square root 
of the integral of 
the square
of the difference of $A$, $B\in {\cal U}$:
\beqa
\Vert A-B\Vert={\sqrt{\frac{1}{2}\int_{M} d^{D}x \;tr 
\;[A^{h}(x)-B^{f}(x)]^{2}}}\; \nonumber
\eeqa
(the extra $\frac{1}{2}$ is included purely for convenience in the
later discussion, and is irrelevant to the arguments in this paragraph). If 
the dimension of space is not equal to 
three, $A$ and $B$ are also defined to be physically equivalent. In 
three space and one time dimension, two connections
$A$ and $B$ are said to be physically equivalent if (\ref{equiv}) holds
and the Chern-Simons index of each of the gauge transformations
$g_{1}$, $g_{2}$,... is equal to zero. 

Both of these new versions of
gauge-equivalence and physical-equivalence are, in fact, equivalence
relations. It is obvious that every connection is equivalent to 
itself. If (\ref{equiv}) holds, then  
$A=\lim B^{g_{n}^{-1}}$. This is because for any $\epsilon>0$, there 
exists a
positive integer $J$, such that if $n>J$, $\epsilon>
\Vert A^{g_{n}}-B \Vert$. But
$\Vert B^{g_{n}^{-1}}-A \Vert=
\Vert A^{g_{n}}-B \Vert$. The 
relations of gauge-equivalence and physical-equivalence are 
symmetric. To prove transitivity of gauge-equivalence, 
suppose that (\ref{equiv}) is valid, and for some connection $C$,  
\beqa
C=\lim B^{h_{n}} \nonumber \;,
\eeqa
for some sequence of gauge transformations $h_{1}$, $h_{2}$,... (which
must, by definition, satisfy (\ref{gt})). Then, defining for each 
$n=1$, $2$,... $f_{n}(x)=g_{n}(x)h_{n}(x)$, one can easily see that each 
$f_{n}$ 
is a gauge transformation (i.e. satisfying (\ref{gt})). The sequence
$A^{f_{n}}$ 
converges to $C$. To see this, note that for any $\epsilon>0$ there exist 
positive integers $J$ and $K$ such that for $n>J$, 
$\Vert A^{g_{n}}-B \Vert <\epsilon/2$ and for $n>K$, 
$\Vert B^{h_{n}} -C\Vert <\epsilon/2$. So if $n$ is larger than both
$J$ and $K$,  
\beqa
\Vert A^{f_{n}}-C \Vert \le 
\Vert A^{f_{n}}-B^{h_{n}} \Vert
+\Vert B^{h_{n}} -C\Vert  
=\Vert A^{g_{n}}-B \Vert
+\Vert B^{h_{n}} -C\Vert<\epsilon\;. \nonumber
\eeqa  

It may seem odd to the reader that two gauge-equivalent connections
are not neccesarily 
transformable to each other by gauge transformations. The reason for
this is the differentiability requirement and the condition 
(\ref{gt}). In fact, if two connections are gauge equivalent
and only one of them is continuous, it is clear that no
``nice" (that is differentiable and square-integrable) gauge 
transformation 
can relate the two.

Let $\a$ and $\b$ be two physical
configurations. Let ${\em G}_{D}={\em G}$ for $D \neq 3$ and
${\em G}_{D}={\em G}_{0}$ for $D=3$. The
distance function on ${\bf \cal M}_{D}$ is defined
by
\beqa
\rho [\a, \b]=  \inf \{ {\sqrt{ I[A,B;h,f]}} \; :\; 
A \in \a, B \in \b,  h\in {\em G}_{D}, f\in {\em G}_{D} 
\} \;, \label{2.1}
\eeqa
where
\beqa
I[A,B;h,f] = \frac{1}{2}
\;\int_{M} d^{D}x \;tr \;[A^{h}(x)-B^{f}(x)]^{2} \label{2.2}
\eeqa
\beqa
= \frac{1}{2}\; \int_{M} d^{D}x 
\;\sum_{k=1}^{D}\; tr[ih(x)^{-1}\pr_{k} h(x)
+ h(x)^{-1}A_{k}(x)h(x)  \nonumber 
\eeqa
\beqa
-\; i f(x)^{-1} \pr_{k} f(x)-f(x)^{-1} B_{k}(x) f(x)]^{2} \;. \nonumber
\eeqa
For the non-practitioner of real analysis, who 
may be discouraged by the notation, the
right-hand-side of equation 
(\ref{2.1}) means ``the greatest
lower bound of the set of
values taken by ${\sqrt{ I[A,B;h,f]}}$ for any
$A \in \a$, $B \in \b$, $h\in {\em G}_{D}$, and 
$f\in {\em G}_{D}$". Since the gauge connections are square 
integrable, the integral (\ref{2.2}) always
converges, but is not
bounded from above. The distance $\rho[\cdot, \cdot]$ can be
either zero or any positive real number.

Since
for any $j \eps {\em G}$
\beq
I[A,B;h,f] = I[A,B;hj,fj]\;, \label{2.200}
\eeq
It is possible to simplify (\ref{2.1}) to
\beq
\rho [\a, \b]=  \inf \{ {\sqrt {I[A,B;h]}} \; \; : \; 
A \in \a ,\; B \in \b , \; h \in {\em G}_{D} 
\}  \;, \label{2.201}
\eeq
where $I[A,B;h] \equiv I[A,B;h,1]$. However, sometimes (\ref{2.1}) is
more convenient to use than (\ref{2.201}), as will be seen below.

The function $\rho [\cdot, \cdot]$ is called
a metric on ${\MM}_{D}$, if for any $\a$, $\b$
and $\sg$ in ${\MM}_{D}$
\beq
\rho [\a, \b ] \ge  0 \;, \label{2.3} 
\eeq
\beq
\rho [\a, \b ] =  \rho [\b, \a ]  \;, \label{2.300} 
\eeq
\beq
\rho [\a, \b ] +   \rho [\b, \sg ] \;\ge\; \rho [\a, \sg ] \;, \label{2.4} 
\eeq
\beq
\rho [\a, \b ]  =  0 \;\; \iff\;\; \a=\b \;. \label{2.5}
\eeq
If these properties are 
satified, then $\MM _{D}$ together with $\rho [\cdot,\cdot]$
is called a metric space. By this 
definition some physical
examples of manifolds for
which
``metrics" are defined
are not really metric spaces (Minkowski space for example). Of 
these four properties, the two which are
most obviously true are (\ref{2.3}) and (\ref{2.300}). Property 
(\ref{2.5}) follows the definition of a Hilbert space
vector as an equivalence class of almost-everywhere-equal functions 
(from
the beginning ``equal" gauge fields were 
defined to be those which are equal 
almost everwhere) and from the definition of physical
equivalence using equation (\ref{equiv}) (in fact the reason
why the equivalence classes were enlarged was precisely so that
(\ref{2.5}) would be satisfied). In order 
to be able to 
prove some very basic analytic properties, e.g., that any convergent
sequence must be a Cauchy sequence, and $\rho [\cdot , \cdot]$ a
continuous function of its arguments, that the space
is Hausdorff, etc. it is
neccesary that the
triangle inequality (\ref{2.4}) be satisfied.

\begin{figure}[t]
\begin{picture}(100,90)(0,0)
\label{1}
\centerline{\epsfbox{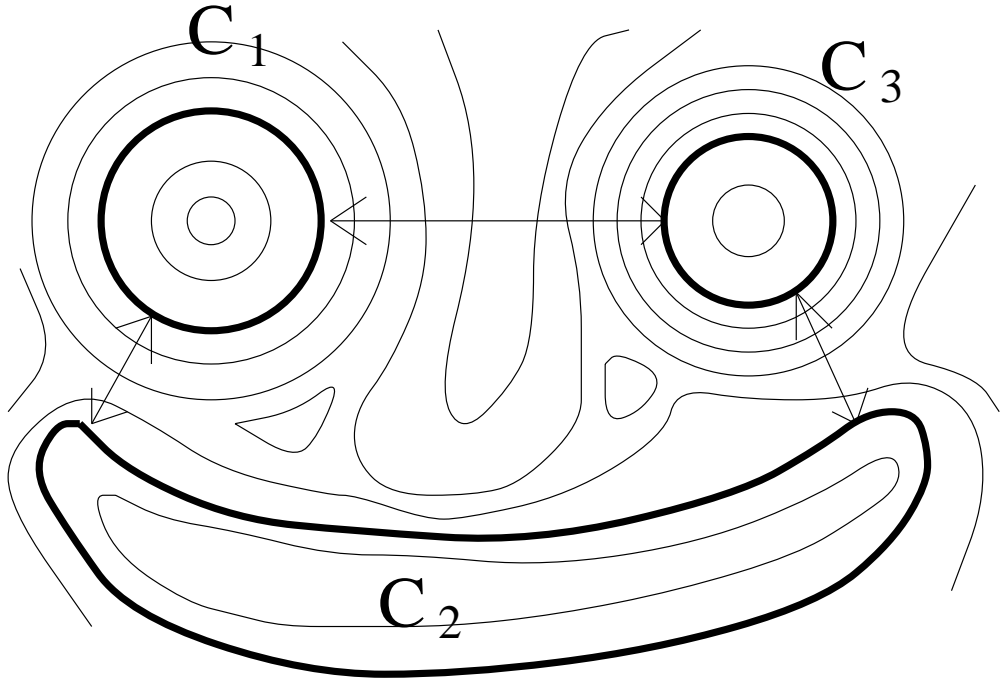}}
\end{picture}
\caption[l]{Consider a family
of curves in the two-dimensional
plane. If the distance between two curves is defined
as the length of the shortest line connecting them, then
this distance does
not always satisfy the triangle inequality. Here 
the distance from curve $C_{1}$ to curve
$C_{2}$ plus the distance
from curve $C_{2}$ to $C_{3}$ is {\em less} than the distance
from curve $C_{1}$ to $C_{3}$.}
\end{figure}

The validity of (\ref{2.4}) is
not entirely trivial. In 
general, coset spaces in Euclidean spaces
do not satisfy the triangle inequality. As an example, consider
a family of curves in $R^{2}$ (see figure 1). If
two points $x, y \in R^{2}$ lie in the same curve, write
$x =_{C} y$. Clearly
$=_{C}$ is an equivalence relation. A 
distance between two equivalence classes (that is, two curves)
can be defined as the minimum of the distance between a point
on one curve and a point on the other. This distance
does {\em not} satify the triangle inequality for 
an arbitrary choice of three elements of the coset space.

\newtheorem{axiom}{Proposition}[section]
\begin{axiom}
: The function $\rho[\cdot, \cdot]$ satifies
the triangle inequality (\ref{2.4}).
\end{axiom}

\noindent
{\em Proof}: Notice that for
any gauge fields $A\in \a$, $B\in \b$ and $S \in \sg$ and
any gauge transformations $h, f, j \eps {\em G}_{D}$
\beqa
{\sqrt{I[A,B;h,f]}}+{\sqrt{I[B,S;f,j]}} \ge {\sqrt{I[A,S;h,j]}} 
\;. \nonumber
\eeqa
Given any real $\epsilon>0$, it is possible
to chose $h$ and $f$ so that ${\sqrt{I[A,B;h,f]}}=\rho [\a,\b]+\epsilon /2$ 
(by the definition of
the greatest lower bound). By property (\ref{2.200}) there
exists a $j$ such 
that ${\sqrt{I[B,S;f,j]}}=\rho [\b, \sg]+\epsilon /2$. Therefore
\beqa
\rho [\a,\b]+\rho [\b,\sg] \ge {\sqrt{I[A,S;h,j]}}-\epsilon \ge 
\rho [\a,\sg] -\epsilon \;.  \nonumber 
\eeqa
Since $\epsilon$ is arbitrary
(\ref{2.4}) follows.

A complete metric
space is one for which any Cauchy sequence, i.e. a sequence
where progressive points get closer to each other, always
has some unique limit point \cite{reed} (the
converse is always true by the
triangle inequality). Thus a complete metric space
has no pathological 
properties, such as isolated points or closed
domains being ``missing" from the space. More
precisely, a sequence 
$\a_{1}$, $\a_{2}$,...,$\a_{n}$,... is a Cauchy sequence if
for any real $\epsilon >0$, there exists
an integer $R$ such that if 
$n,m >R$, then $\rho[\a_{n}, \a_{m}]< \epsilon$.

\begin{axiom}
: The metric space ${\MM}_D$ is complete.
\end{axiom}

\noindent
{\em Proof}: Recall the 
metric on the space of gauge 
connections, ${\cal U}$ given above. For $A$, $B\in {\cal U}$, this is
\beqa
\Vert A-B\Vert={\sqrt{I[A,B;1,1]}}\;. \nonumber
\eeqa
With this metric, ${\cal U}$ is 
a complete metric space (in fact, a Hilbert space, each whose
vectors is representated by a set of 
measurable, square-integrable, Lie-algebra-valued functions, all 
equal except on a set of measure zero). Let 
$\a_{1}$, $\a_{2}$,...,$\a_{n}$,... be
a Cauchy sequence. Then for any $\epsilon >0$ there exists
some $R$ such that for $n,m>R$, $\rho[\a_{n}, \a_{m}]< 
\epsilon/2$. From the definition of the metric and
from property (\ref{2.200}) it is
possible to find 
$A_{1} \in \a_{1}$, $A_{2} \in \a_{2}$,..., $A_{k} \in \a_{k}$, such
that for $n,m>R$
$\Vert A{n}-A_{m} \Vert = 
\rho[\a_{n},a_{m}] + \epsilon/2 <\epsilon$. Thus, since 
$\Vert \cdot - \cdot \Vert$
is a metric on a complete metric space ${\cal U}$, the sequence 
$A_{1}$, $A_{2}$,...,$A_{n}$,... converges in this metric. Since
$\rho[\a,\b]$ is bounded above by $\Vert A-B \Vert$, the sequence
$\a_{1}$, $\a_{2}$,...,$\a_{n}$,... therefore 
converges in ${\MM}_{D}$.

%By standard methods in the calculus of variations, the lower
%bound in (\ref{2.1}) is in fact saturated. This is actually a 
%consequence of the completeness of the metric $\Vert \cdot -\cdot \Vert$
%in $\cal U$. Consequently, the
%infimum can be replaced by the absolute minimum:
%\beq
%\rho [\a, \b]=  \min \{ {\sqrt{ I[A,B;h,f]}} \; :\; 
%A \in \a, B \in \b,  h\in {\em G}_{D}, f\in {\em G}_{D} 
%\} \;. \label{2.10}
%\eeq
%There is no reason to expect
%that for a given $A$ in $\a$ the choice
%of $B$ in $\b$ which attains this minimum will be 
%unique. 

Though complete, the metric space ${\MM}_{D}$ is not compact, for 
it is not totally bounded or
even bounded. It is
possible to make 
the space bounded in a variety of ways. For example, one can define 
\beqa
\rho_{\L} [\a, \b]=  \inf \{ {\sqrt{ I_{\L} [A,B;h,f]}} \; :\; 
A \in \a, B \in \b,  h\in {\em G}_{D}, f\in {\em G}_{D} 
\} \;, \nonumber
\eeqa
where
\beqa
I_{\L}[A,B;h,f] = \frac{1}{2}\; \int_{M} d^{D}x 
\; \frac{tr[A^{h}-B^{f}]^{2}}
{\{ 1+ \frac{1}{\Lambda} {\sqrt {tr[A^{h}-B^{f}]^{2}}}\; \}^{2} 
}\;, \nonumber
\eeqa
It can be shown with some work that this distance function is a 
metric and that
the metric space so defined is
complete. The reader will notice that there is an upper bound on this
distance: $\rho_{V,\L}[\a,\b] \le V \L^{2}/2$. However, while
this metric space is bounded, it is not obviously totally bounded. I do
not have a proof that the space is not totally bounded, but this
possibility seems rather remote. This is unfortunate, for a
space of compact closure
is needed to prove many of the standard results on the spectra
of coordinate-invariant Laplace operators \cite{davies} (I don't know
which of 
these results are generally
true without assuming compact closure). In fact, the
only way I know of to make the space of gauge configurations
compact is to use a lattice, which will be discussed in a later
publication \cite{kudinov}. A distance
function analogous
to $\rho[\cdot, \cdot]$ can be defined on a
lattice. When the
lattice is finite, this has the virtue
of making ${\MM}_{D}$ a compact manifold. Futhermore, the metric
tensor is still idempotent, making it possible to define
the generalized inverse (\ref{1.5}).

It is not easy to calculate the distance between two
arbitrary physical configurations. In particular, the variational
problem of finding the gauge transformations minimizing
$I[A, B;h,f]$ is a nonlinear set of differential equations. In 
section 7, it is shown that for $D=2$ and 
$B$ a pure 
gauge, these
equations are completely integrable.

\section{Geodesics and the intrinsic metric
in ${\MM}_{D}$}

The
geodesics on ${\MM}_{D}$ were
found by Babelon and Viallet \cite{babelon2} using the tangent-space
inner
product of Singer \cite{singer}. They
also conjectured
that the distance function $\rho[\a,\b]$ 
is the geodesic distance function between two
configurations $\a$ and $\b$ in ${\MM}_{D}$. What 
is interesting is that even
before doing any Riemannian geometry (which takes some justification! See
section 6) the geodesics can be easily
found from the definition of the metric $\rho[\cdot, \cdot]$ in equations 
(\ref{2.1}) and (\ref{2.2}) using unsophisticated
methods. Furthermore, Babelon and Viallet's conjecture is simple to
verify.

In any metric space, there is a second 
metric, called the {\bf intrinsic} 
metric $r[\cdot, \cdot]$ defined as the minimal length of a 
curve (parametrized by a real number)
connecting two 
points \cite{alex}. It is
not hard to prove that $r[\cdot, \cdot]$ satisfies
all the conditions
of a metric space. It is easy to show that
for any two configurations 
$\a$ and $\b$, $r[\a,\b] \ge \rho[\a,\b]$. Furthermore, the two metrics
are asymptotically equal as the two points coalesce.

What will be shown in this section is that not 
only is the intrinsic
metric $r[\cdot,\cdot]$
bounded above by $\rho[\cdot,\cdot]$, but the
two metrics are actually the same (as conjectured
in reference \cite{babelon2}). Thus the distance defined
in section 4 
between two gauge configurations is the length of a minimal
curve joining them together. This 
curve corresponds to a particular straight line
in ${\cal U}$ (though it is certainly not straight in ${\MM}_{D}$).

Only minimal geodesics bewteen
pairs of points will be considered in this section.

The intrinsic metric 
between two points, $\a$
and $\a^{\r}$, is obtained by \begin{enumerate}
\item picking a real number $\epsilon >0$,
\item considering sets of $n$ points, 
$\a_{1}$,...,$\a_{n} \in {\MM}_{D}$, such
that $\rho[\a, \a_{1}]$, $\rho[\a_{1}, \a_{2}]$,...$\rho[\a_{n}, \a^{\r}]$
are all less than $\epsilon$,
\item finding the greatest lower bound of the sum of distances
$\rho [\a, \a_{1}]+\rho [\a_{1}, \a_{2}]+...+\rho [\a_{n}, \a^{\r}]$
with respect to $\a_{1}$,...,$\a_{n}$,  
\item finding the greatest lower bound of this sum
with respect to the number of interpolating points
$n$ and 
\item taking the limit as $\epsilon$ goes to zero.
\end{enumerate}
This definition can be shown to be equivalent to that
of Aleksandroff and Zalgaller 
\cite {alex}. The number of 
interpolating points, $n$, generally (but
not always) becomes infinite. If 
a parameter $t \in [0,1]$ is introduced by
$\a_{j}=\a(t)$, with $t=j/n$, the points form a minimal curve.

Let $A^{\r}$ be a connection in $\a^{r}$ Let 
${\MM}_{D}^{\epsilon,n}$ be the set of $n$-tuples $(\a_{1},...\a_{n})$
in ${\MM}_{D}$ such that 
$\rho[\a, \a_{1}] < \epsilon$, $\rho[\a_{1}, \a_{2}] < \epsilon$... 
$\rho[\a_{n}, \a^{r}] < \epsilon$. Let 
${{\cal U}_{h, A^{\r}}}^{\epsilon, n}$
be the set of $n$-tuples $(A_{1},...A_{n})$ in ${\cal U}$ such that 
$\Vert A^{h} - A_{1} \Vert < \epsilon$, $\Vert A_{1} -A_{2} 
\Vert < \epsilon$,..., $\Vert A_{n}-A^{\r} \Vert < \epsilon$. The 
definition of the intrinsic metric is
\beqa
r[\a,\a^{\r}]
&=& \lim_{\epsilon \rightarrow 0}\;\; 
    \inf_{n=0,1,2...}\;\; \inf_{(\a_{1},...\a_{n}) \in 
    {{\MM}_{D}}^{\epsilon,n}}
    \{ \rho[\a,\a_{1}]+ \rho[\a_{1},\a_{2}] \nonumber \\
&+& \cdot \cdot \cdot +
\rho[\a_{n-1},\a_{n}]+ \rho[\a_{n},\a^{\r}] \} \;. \nonumber
\eeqa
By virtue of property (\ref{2.200}), this is equal to
\beqa
r[\a,\a^{\r}]
&=& \lim_{\epsilon \rightarrow 0}\;\;
    \inf_{n=0,1,2...} \;\;
    \inf_{h,h_{1},...,h_{n} \in {\em G}_{D}}\;\; 
    \inf_{A\in \a, A^{\r}\in \a^{\r}}\;\;
    \inf_{({A_{1}}^{h_{1}},
      ...,{A_{n}}^{h_{n}}) \in {{\cal U}_{h, A^{\r}}}^{\epsilon, n} }\;
     \{ {\sqrt{I[A,A_{1};h,h_{1}] }}  \nonumber \\
&+& {\sqrt{I[A_{1},A_{2};h_{1},h_{2}] }}
+ \cdot \cdot \cdot + {\sqrt{I[A_{n-1},A_{n};h_{n-1},h_{n}] }}
+ {\sqrt{I[A_{n},A^{\prime};h_{n},1] }}     \; \}\;. \nonumber 
\eeqa
or, equivalently
\beqa
r[\a,\a^{\r}]
&=& \lim_{\epsilon \rightarrow 0}\;\;
    \inf_{n=0,1,2...}    \;\;
    \inf_{h,h_{1},...,h_{n} \in {\em G}_{D}}\;\;
     \inf_{A\in \a, A^{\r}\in \a^{\r}, (A_{1}^{h_{1}},...,A_{n}^{h_{n}}) 
     \in {{\cal U}_{h, A^{\r}}}^{\epsilon, n}  }\;    
     \{ \Vert A^{h}-A_{1}^{h_{1}} \Vert         \nonumber \\
&+& \Vert A_{1}^{h_{1}}-A_{2}^{h_{2}} \Vert
+ \cdot \cdot \cdot
\Vert A_{n-1}^{h_{n-1}}-A_{n}^{h_{n}} \Vert    
+ \Vert A_{n}^{h_{n}}-A^{\prime} \Vert   \}  \;. \label{g.3}
\eeqa
The connections $A_{j}^{h_{j}}$ in (\ref{g.3})
can be replaced by just $A_{j}$, which
gives
\beqa
r[\a,\a^{\r}]
&=& \lim_{\epsilon \rightarrow 0}\;\;
    \inf_{n=0,1,2...} \;\;
    \inf_{h \in {\em G}_{D}}\;\;
    \inf_{A\in \a, A^{\r} \in \a^{\r} }\;\;
    \inf_{ A_{1}, ..., A_{n} \in {{\cal U}_{h,A^{\r}}}^{\epsilon, n}     } \;
    \{ \Vert A^{h}-A_{1} \Vert 
    + \Vert A_{1} -A_{2} \Vert   \nonumber \\
&+& \cdot \cdot \cdot
+\Vert A_{n-1} -A_{n} \Vert    
+ \Vert A_{n} -A^{\prime} \Vert   \}  \;. \label{g.4}
\eeqa
Since ${\cal U}$ is a linear space, the infimum over $A_{1}$,...,$A_{n}$
of the quantity in brackets $\{ \}$ in
(\ref{g.4}) is just $\Vert A^{h} - A^{\r} \Vert$. Thus
(\ref{g.4}) reduces to
\beqa
r[\a,\a^{\r}]=\rho[\a,\a^{\r}] \;, \nonumber
\eeqa
which is what was conjectured by 
Babelon and Viallet \cite{babelon2}.

It is now obvious what the geodesic joining $\a$ and $\b$ is. Let
$\Xi$ be the mapping from ${\cal U}$ to ${\MM}_{D}$ which takes
any gauge field configuration into the physical configuration
containing that gauge field 
configuration (for example, $\Xi (A)=\a$). If
$A \in \a$ then there is 
a choice of $B \in \b$, which minimizes
$\Vert A-B \Vert$. The geodesic
is the image under $\Xi$ 
of the line segment in ${\cal U}$
\beqa
A(t)=A+Et \;, \label{d.145} 
\eeqa
where $A+E=B$ and
$0\le t \le 1$. This
solution was found using different
methods by Babelon and Viallet \cite{babelon2}. It
is just a straight line in connection space ${\cal U}$. However, the
geodesic is {\em not} a
straight line in ${\MM}_{D}$. The geodesic curve is a curve in
the space of equivalence classes, $\Xi(A(t))$, and not a curve in
the space of
connections, $A(t)$. Notice
that because the geodesic is parametrized by a straight 
line in connection
space, ${\cal U}$, it cannot
contain a point conjugate to the point $\Xi(A)$ at $t=0$. In
fact, one of the technical 
inconveniences of the formalism
used here is that conjugate points and closed geodesics
cannot be obtained from just one solution (\ref{d.145}). They
must be made
by more than one such solution (\ref{d.145}) pieced together at the ends
by gauge 
transformations.

Other
extremal curves, conjugate points and closed geodesics can probably
be examined using the methods developed in this paper. It seems
reasonable that the curve $\Xi(A(t))$ is still a geodesic for
$t<0$ and $t>1$. Does extending the geodesic
in this way always make a
closed curve? If so, there are an infinite number of Gribov
copies of $A$ along the line $A(t)\in {\cal U}$ \cite{gribov}.

\section{The singular metric tensor on ${\MM}_{D}$}

Rather picking out members
of each equivalence class of gauge fields $\a$ by
than gauge-fixing, the philosophy advocated in this paper is
to work directly
with the space of equivalence classes. A straightforward way
to do this is to use connections in $\cal U$ as the coordinates
of equivalence classes in ${\MM}_{D}$. The advantage
of such a procedure is that the Gribov problem of
copies
is never confronted \cite{gribov}. The
disadvantage is that the metric tensor is 
singular. Here the word ``singular" 
means that some eigenvalues of the metric tensor
are zero; it does not mean that
the metric is discontinuous
or divergent at some point of configuration space ${\MM}_{D}$. In
spite of this singularity, it
is possible to define standard geometric quantities 
on ${\MM}_{D}$, such as the
the volume element, the Laplacian and the Riemann curvature
tensor. Since differential calculus with
singular metric tensors does not seem to be discussed in 
textbooks on geometry, a review is given in the appendix.

To obtain the metric tensor, one would like
to consider two ``close" physical
configurations $\a$, $\b$ containing gauge configurations $A$
and $A+\delta A$, respectively, where $\delta A$ is 
infinitesmal, thereafter expanding to quadratic
order in $\delta A$. If this procedure 
is to make sense, certain criteria must
be satisfied:
\begin{enumerate}
\item If $\b$ 
is some configuration near 
$\a$ and $\a$ contains $A$, then $\b$
contains a gauge field 
$B$ of the form $A+\delta A$. 
\item The connections in ${\cal U}$ are mapped continuously
to physical configurations in ${\MM}_{D}$. 
\item If $\b$ contains two distinct
elements $B=A+\delta A$ and $C=A+(\delta A)^{\r}$ then it must
be so that $B$ and $C$ differ by a ``small" gauge transformation
${\cal D}^{A} \omega \approx (\delta A)^{\r}-\delta A$. 
\end{enumerate}

Criteria 1 and 2 are needed in order to be able to use
connection space, $\cal U$, as coordinates for physical
configuration space, ${\MM}_{D}$. Criterion 3 is required if
the metric is to be Taylor expanded in these coordinates
to obtain the metric tensor and its derivatives. One might
worry that criterion 3 is
false; in particular that $C$ could be a Gribov copy of $B$. This would
mean that $B$
and $C$ are 
equivalent to each other
by a ``large" gauge transformation $g$ (of
Chern-Simons number zero for $D=3$), $B^{g}=C$, but
are not equivalent by
a ``small" gauge transformation, $\exp-i\omega(x)$, for which
${\cal D}^{A} \omega \approx C-B$. Fortunately, all
the criteria are true, as
will be shown below.

Before 
proving these criteria, a few
definitions will be made. Recall from section 4 that
${\cal U}$ is
a metric space, with the metric $\Vert \cdot -\cdot \Vert$. Recall
from section 5 that 
$\Xi$ is the mapping from ${\cal U}$ to ${\MM}_{D}$ which takes
any gauge field configuration into the physical configuration
containing that gauge field 
configuration. The open
ball $s_{r}(\a)$ of radius $r$ around $\a$ is defined by
\beqa
s_{r}(\a)=\{ \g : \rho [\g,\a] <r \}\;. \nonumber
\eeqa
and the open ball $S_{r}(A)$ of radius $r$ around $A$ is defined
by
\beqa
S_{r}(A)=\{ C : \Vert A-C \Vert  <r \}\;. \nonumber
\eeqa

Criteria 1, 2 and 3, respectively, can now be stated as:

\begin{axiom}
: Suppose 
$A \in {\cal U}$ is any gauge field configuration 
and $\Xi (A)=\a$. Then 
$s_{r}(\a)$ is the image under $\Xi$ of $S_{r}(A)$, or
\beq
\Xi(S_{r}(A))=s_{r}(\a)\; .  \label{3.7}
\eeq
\end{axiom}

\noindent
{\em Proof:} By the definition 
of $\rho[\cdot, \cdot]$ and $\Vert \;\Vert$
for any $B \in S_{r}(A)$,
\beqa
\rho[\a, \Xi(B)]=\rho[\Xi(A), \Xi(B)] \le 
{\sqrt{I[A,B;1,1]}}=\Vert A-B\Vert \;. \nonumber
\eeqa
Therefore $\Xi(S_{r}(A))$is contained
in $s_{r}(\a)$. Now 
suppose $\b$ is in $s_{r}(\a)$. That means, that
there is a $B\in 
\b$ such that  
there is sequence of gauge transformations $h_{1}$, $h_{2}$,..., with
$\Vert \lim B^{h_{n}}-A \Vert <r$. But then, for sufficiently
large $n$, $\Vert B^{h_{n}}-A \Vert <r$.  Therefore, since
$B^{h_{n}} \in \b$ and 
$B^{h_{n}} \in S_{r}(A)$, $\Xi(S_{r}(A))$ is 
contained in $s_{r}(\a)$. Equation (\ref{3.7}) follows.

\begin{axiom}
: The mapping $\Xi$ is continuous.
\end{axiom}

\noindent
{\em Proof:} In fact, a stronger condition can be proved, namely
that $\Xi$ is uniformly continuous. This means that for
any $\epsilon >0$ there exists a $\delta >0$ such that if
$\Vert A-B\Vert <\delta$ then $\rho[\Xi(A), \Xi(B)] <\epsilon$. For
the case at hand, it is obvious that one can choose $\delta=\epsilon$.

%Consider any point $\a$ in ${\MM}_{D}$ and
%the open ball $s_{r}(\a)$. Let $A$ be some gauge field
%in ${\cal U}$, such that $\Xi(A)=\a$. Then by 
%Proposition 1 the inverse image $\Xi^{-1}[s_{r}(\a)]$
%is the union of 
%$S_{r}(A^{h})$ for all gauge transformations $h$:
%\beqa
%\Xi^{-1}[s_{r}(\a)] = \cup_{h} S_{r}(A^{h})\;.
%\eeqa
%Since $\Xi^{-1}[s_{r}(\a)]$
%is a union of open sets, it must itself be open. Therefore
%$\Xi$ is continuous.

\begin{axiom}
: Suppose $B,C \in S_{r}(A)$ 
and that $\Xi(B)=\Xi(C)=\b$. For $r$ sufficiently small, there exists a
gauge transformation $g(x) \in SU(N)$ such that
\beq
C_{k}(x)=g(x)^{-1} B_{k}(x)g(x)
+ig(x)^{-1} \partial_{k}g(x) \;, \label{3.8}
\eeq
and such that $g$ can be expanded in the form:
\beqa
g(x)=1-i \int_{M} d^{D}x \sum_{k=1}^{D}  
S_{k}(x) [C_{k}(x)-B_{k}(x)] T_{k}(x) \nonumber
\eeqa
\beqa
-\int_{M} d^{D}x \int_{M} d^{D}y \sum_{k=1}^{D} \sum_{l=1}^{D}
U_{k,l}(x, y) [C_{k}(x)-B_{k}(x)]              \nonumber
\eeqa
\beq
\times V_{k,l}(x, y)[C_{l}(y)-B_{l}(y)]
W_{k,l}(x, y)      
+O(r)^{3} \;,  \label{3.9}
\eeq
where $S_{k}(x)$, $T_{k}(x)$, $U_{k,l}(x, y)$, $V_{k,l}(x, y)$ 
and $W_{k,l}(x, y)$                 
are all matrix valued generalized functions.
\end{axiom}

\noindent
{\em Proof:} Equation (\ref{3.8}) can be solved explicitly for 
$g(x)$, given that $B$ and $C$ are gauge-equivalent. For
each point, $x \in M$, define $l$ to be the straight-line path from 
the origin to $x$. Consider the expression 
\beq
g(x)=[{\cal P} \exp -i\int_{0}^{x} dz \cdot B(z)] \;g(0)\;
[{\cal P} \exp -i\int_{0}^{x} dz \cdot C(z)]^{\dag} \;, \label{3.10}
\eeq
where both integrations are along $l$ and
where $g(0) \in SU(N)$ is an integration constant. Since
any solution of the form (\ref{3.10}) formally
solves (\ref{3.8}), $g(0)$
can be chosen to be unity. The task is then to show
that (\ref{3.10}) really exists as a gauge transformation. Now
(\ref{3.10}) 
can be functionally expanded in $C-B$ as
\beqa
g(x)=1 
&   +   & i\sum_{k} \int_{0}^{x} dy^{k}[{\cal P} 
         \exp -i\int_{y}^{x} dz \cdot B(z)] 
         [C_{k}(y)-B_{k}(y)]                \nonumber \\
&\times & [{\cal P} \exp -i\int_{y}^{x} dz \cdot C(z)]^{\dag}  \nonumber  
\eeqa
\beq
+(quadratic\;\; term\;\; in\;\; C-B)+O(C-B)^{3}\;. \label{3.11}
\eeq
Since $\Vert C-B \Vert < r$, making $r$ sufficiently
small (recall that $M$ is compact) guarantees that this expansion 
converges and that it may be differentiated term-by-term
with respect to $x$. One can also check that it satisfies
(\ref{gt}). Note that equation 
(\ref{3.11}) is of the form (\ref{3.9}). This concludes the proof.

The rigorous part of this paper is now finished. What remains in 
this section could presumably be made rigorous by being careful
with domains of covariant derivative operators on the Hilbert
space ${\cal U}$.

Now that the use of gauge fields as local coordinates on the space
of physical configurations
is justified, the metric tensor can be
obtained by means of a functional
Taylor expansion $g \approx 1-i\omega$. Suppose
that the distance
between $\a$ and $\b$ in ${\MM}_{D}$
is a small quantity $\rho [\a, \b]=d\rho$, and the distance
in ${\cal U}$ between $A$ in $\a$ and $B$ 
in $\b$, namely $\Vert B-A \Vert$, is 
also small. Then
if $\delta A$ is defined as $B-A$,
\beq
d\rho ^{2}=
\min \{ \; \frac{1}{2}\;\int_{M} d^{D}x \;\,tr\;\sum_{j=1}^{D}
[A^{g}_{j}-A_{j}-\delta A_{j}]^{2}\;\; : 
\;g \in {\em G}_{D} \} \;. \label{4.1}
\eeq
In (\ref{4.1}), Proposition 6.3 has been use to replace the infimum 
by the minimum. The variational condition on $g \approx 1-i\omega$ is
\beq
{\cal D}^{j}{\cal D}_{j} \omega = {\cal D}^{j}\delta A_{j} \;. \label{4.2}
\eeq
The solution of (\ref{4.2}) exists by virtue of the arguments
given at the end of section 4. It is 
\beqa
\omega ={\cal P}
\frac{1}{{\cal D}^{2}} \;{\cal D}^{j}\delta A_{j} \;, \label{4.200}
\eeqa
where in the Green's function ${\cal P}\frac{1}{{\cal D}^{2}}$ taking
the
principle value projects out the zero
modes of ${\cal D}^{2}$ according to the prescription (\ref{1.5}) in 
section 3. Thus the problem
with non-generic
points raised in reference \cite{babelon2} never
arises. In fact, it is clear from section 4
that such a problem shouldn't
occur since ${\MM}_{D}$ is a 
complete metric space. Substituting (\ref{4.200})
back 
into (\ref{4.1}) then gives 
\beqa
d\rho ^{2}=
[\int_{M} d^{D}x \;\sum_{j=1}^{D} \;\sum_{a=1}^{N^{2}-1}]
\;[\int_{M} d^{D}y \;\sum_{k=1}^{D}\;
 \sum_{b=1}^{N^{2}-1}]\;
G_{(x, j, a) \;(y, k, b)}\; \delta A^{a}_{j}(x) 
\delta A^{b}_{k}(y)\;, \nonumber
\eeqa
where the metric tensor is
\beq
G_{(x, j, a) \;(y, k, b)}=\delta_{j\;k} \delta_{a\;b} \delta^{D}(x-y)
-({\cal D}_{j}\; {\cal P}\frac{1}{{\cal D}^{2}} \;{\cal D}_{k})_{a\;b}\;
\delta^{D}(x-y)\;, \label{4.4}
\eeq 
which is (\ref{1.4}). The metric tensor clearly
defines a symmetric form. It is probably possible (being careful
in defining some dense subspace of Hilbert space) to define it as
a sef-adjoint operator. As stated 
earlier, $G$ is the projection operator onto small
variations of $A$ which are not gauge
transformations. Therefore $G$ is singular. This singularity
is the price to pay for not fixing the gauge. 

As already pointed out in section 3, the metric tensor
should be intepreted as its own inverse. This tensor
needs to be regularized in the kinetic term of
the Yang-Mills Hamiltonian.

The analysis above
can be repeated for an Abelian gauge theory. The
result for the
metric tensor is not very different from (\ref{4.4}):
\beqa
G_{(x, j) \;(y, k)}=\delta_{j\;k}  \delta^{D}(x-y)
-\pr _{j}\; {\cal P}\frac{1}{\pr ^{2}} \;\pr _{k}\;
\delta^{D}(x-y)\;. \nonumber
\eeqa
Since this metric tensor is independent of
the gauge field, ${\MM}_{D}$ is 
flat. The metric is easy to calculate for arbitrarily
(not infinitesmally) seperated Abelian gauge fields $A_{k}(x)$ 
and $B_{k}(x)$ in equivalence classes $\a$ and 
$\b$, respectively. For $D=2$, it is
\beqa
\rho[\a, \b]=-\int d^{2}x \int d^{2}y\; 
[{\bf \nabla} {\bf \times} ({\bf A(x)}-{\bf B(x)})] \, \frac{1}{2\pi}\,
\ln \frac{|x-y|}{a} \,                          
[{\bf \nabla} {\bf \times} ({\bf A(y)}-{\bf B(y)})] \;, \nonumber 
\eeqa
where $a$ has dimensions of length, and for $D=3$ it is
\beqa
\rho[\a, \b]=\int d^{3}x \int d^{3}y\; 
[{\bf \nabla} {\bf \times} ({\bf A(x)}-{\bf B(x)})] 
\cdot \frac{1}{4\pi|x-y|}\,\,
[{\bf \nabla} {\bf \times} ({\bf A(y)}-{\bf B(y)})] \;. \nonumber
\eeqa

The measure of integration on ${\MM}_{D}$ needs to be defined 
to make
the space of states of the Yang-Mills theory
meaningful. Formally, if
two wave functionals are $\Psi[\a]$ and $\Phi[\a]$ and $Q$ is some
operator, the matrix element is the functional integral
\beqa
<\Psi \vert Q \vert \Phi>=\frac{\int dA\; {\sqrt{det^{\prime} G}}\;
\Psi[\Xi(A)]^{*} \;Q\; \Phi[\Xi(A)]}
{\int dA\; {\sqrt{det^{\prime} G}}\;
\Psi[\Xi(A)]^{*}\; \Phi[\Xi(A)]}\;, \label{3.66}
\eeqa
where ${\sqrt{det^{\prime} G}}$ is the determinant of the metric
tensor after the removal of zero modes. The determinant
${\sqrt{det^{\prime} G}}$ is unity in the unregulated
theory (it is still unity with lattice regularization 
\cite{kudinov}). However, the integrals in the numerator
and denominator of (\ref{3.66}) are divergent unless a regulator
is introduced. 

There
is no difficulty with the expression (\ref{3.66}) if the volume
of each gauge orbit $O_{\a}=\{ A \in {\cal U} \; : \Xi(A)=\a \}$
is independent of $\a$. However, this
is not true everywhere on ${\MM}_{D}$. At certain 
points of ${\MM}_{D}$, called 
reducible connections in reference \cite{fuchs}, there is
a nontrivial invariant subgroup of gauge transformations
which preserve the gauge configuration. At these points, the
gauge orbit
has zero volume (if the invariant
subgroup is infinite) or a nonzero, but reduced
volume (if the invariant subgroup is finite. The volume
must be divided by the order of this subgroup). At a reducible
connection $A$, there is a special-unitary-matrix-valued 
function $g(x) \in SU(N)$ which is an eigenfunction of ${\cal D}^{2}$, with
unit eigenvalue, i.e.
\beqa
{\cal D}^{2} g=g \;. \nonumber
\eeqa
Under a small perturbation of $A$, such a property is
not generally
maintained. Therefore (while I have not studied this question
carefully) the reducible connections are a set of
measure zero in the integral (\ref{3.66}) and should
present no fundamental difficulty.

\section{An integrable system related to the metric of 
$2+1$-dimensional gauge theory}

In section 4 the distance between arbitrary
two Yang-Mills 
configurations in ${\MM}_{D}$
was defined and proven to be a metric. Before 
pursuing the infinitesmal form
of the metric further, a good question is: how 
close one can come to actually
calculating the metric $\rho[\cdot,\cdot]$ for
two arbitrary points in
${\MM}_{D}$? The answer, at least
for the time being is: not very. However, the
problem has some
interesting aspects which may eventually point toward
a solution.

The first step towards a general
calculation is to 
minimize $I[A, B ; h]$. If this functional is minimized, then 
for $N^{2}-1$ real numbers, $\omega^{a}(x)$ at each point $x$.
\beq
\frac{\del 
I[A, B ; he^{i\omega^{a}t_{a}}]}{\del \omega^{a}(x)}=0 \;. \label{3.1}
\eeq
For the case of $B$ equal to
a pure gauge, this problem has been considered before, because its solution
would make it possible to locate the
fundamental region \cite{zwanziger, s-t-s}.

The variational action principle is that of a classical chiral sigma
model, implying the differential equations in $D$ dimensions
\beq
\pr^{j}[(A^{h}_{j}(x))^{a}-B_{j}^{a}(x)]-
f_{a\;b\;c}(A^{h}_{j}(x))^{b}B_{j}^{c}(x) =0\;, \label{3.2}
\eeq
where 
\beq
A^{h}_{j}(x)=ih(x)^{-1} \pr_{j} h(x)+ 
h(x)^{-1} A_{j}(x) h(x)\;.  \label{3.3}
\eeq
The usual chiral sigma model, described by (\ref{3.2}) with
both $A_{k}(x)=0$ and $B_{k}(x)=0$, is well known to be integrable with
an infinite set of conserved charges \cite{poly}.

The condition (\ref{3.1}) only guarantees that $I[A,B;h]$ is an
extremum, which may not even be a local minimum. Notice that 
$I[A,B;h]$ is
the energy of a gauged spin model with a fixed external gauge field; in
other words a spin glass. The specific solution needed for 
calculating $\rho[\a,\b]$ is 
the {\em absolute} minimum of 
$I[A, B; h]$ \cite{zwanziger, s-t-s}. The 
absolute minimum of the energy
of a spin glass
is a notoriously difficult quantity to determine 
\cite{spingl}.

Notice that if $B=0$, (\ref{3.1}) 
becomes
\beq
\pr^{j}[(A^{h}_{j}(x)]^{a}=0\;, \label{3.2222}
\eeq
which is the Coulomb gauge 
condition \cite{zwanziger, van baal, cutkosky}. The 
solution of
(\ref{3.2222}) is certainly
not unique \cite{gribov}. Nonetheless, the
analysis presented in this section
seems worthwhile; eventually it may help to
shed light on the nature of the absolute minimum.

In $2+1$ dimensions, the two-dimensional nonlinear
equations (\ref{3.2}) and (\ref{3.3})
are completely integrable when $B_{j}(x)=0$; that
is they may be written as the compatibility conditions
of a two sets of linear equations. This means that
the these equations may be expressed as the
vanishing of the commutator of two operators, called
a Lax pair. A reason why it would be useful to
obtain the solution of these equations 
with $B_{j}(x)=0$, in order to find
$\rho [\a,0]$, where $0$ is the equivalence
class of pure gauge configurations, is that this quantity
yields a bound on the distance between any two 
physical
configurations $\a$ and $\b$ in ${\MM}_{2}$, by the 
triangle inquality (\ref{2.4})
\beqa
\rho [\a, 0 ] +   \rho [\b, 0 ] \;\ge\; \rho [\a, \b ] \;. \nonumber
\eeqa
Furthermore, it happens that the value of the
metric when one of the points is a pure gauge configuration is
one of the most physically interesting cases, as will be discussed in
sections 10 and 11. 

There is 
another way of 
writing (\ref{3.2}) and (\ref{3.3}) for $h$
when $B_{j}(x)=0$. 
Define the curvature free gauge field $H$ by 
\beqa
H_{j}(x)=ih(x)^{-1} \pr_{j} h(x) \;, \label{3.333}
\eeqa
so that for any contour from a boundary point $y$ to the point $x$
\beqa
h(x)= {\cal P} \exp -i \int_{y}^{x} dz \cdot H(z) \equiv
V_{y}^{x} [H] \;. \nonumber
\eeqa
The equations (\ref{3.2}), (\ref{3.3}) and (\ref{3.333}) are therefore the
same as
\beqa
[\pr_{1}-iH_{1}(x),\pr_{2}-iH_{2}(x)]
&=&0 \;, \nonumber \\
    \pr^{k} \{ H_{k}(x)+ (V_{y}^{x} [H])^{\dag} A_{k}(x) V_{y}^{x} [H] \} 
&=&0\;. \label{3.4}
\eeqa

Define
$G(x;[A],[H])$ by
\beq
G(x;[A],[H])=\frac{1}{\pr_{1}-i[H_{1}(x),\cdot]+\lambda \pr_{2}} \;
\pr^{k} \{ (V_{y}^{x} [H])^{\dag} A_{k}(x) V_{y}^{x} [H] \}
\;,   \label{3.6}
\eeq
with boundary condition $G(x;[A], [H])\rightarrow 0$ as $x^{1}\rightarrow
\pm \infty$. It is not
generally possible to integrate (\ref{3.6}) 
in closed
form for $G(x;[A],[H])$, but this 
quantity can be defined
as a formal power series. Equation (\ref{3.4}) is equivalent to
\beq
[\pr_{1}-iH_{1}(x)-\lambda \pr_{2},\,\, 
\pr_{2}-iH_{2}(x)+\lambda \pr_{1}-i\lambda G(x;[A],[H])]=0\;, \label{3.5}
\eeq
for arbitrary parameter $\lambda$. The two operators
in the commutator in the left-hand-side of 
(\ref{3.5}) is the Lax pair.

If the coordinate $x^{2}$ is identified with the time and 
if the gauge field $H_{k}(x)$ falls to zero as 
$x^{1} \rightarrow \pm \infty$, equation
(\ref{3.5}) can be used to derive an infinite tower of conserved 
charges \cite{poly}. Define the connection $K_{j}(x) \in su(N)$
by
\beqa
K_{1}=H_{1}\;,\;\; K_{2}=H_{2}+G(x;[A],[H])\;. \label{3.61}
\eeqa
Upon substitution
of (\ref{3.61}), the equations of motion, (\ref{3.5}), take the form
\beqa
[\pr_{1}-iK_{1}(x)-\lambda \pr_{2},\,\, 
\pr_{2}-iK_{2}(x)+\lambda \pr_{1}]=0\;.  \nonumber
\eeqa
By making a rotation of angle $\theta=\tan^{-1} \lambda$, (\ref{3.8}) becomes
\beqa
[\pr_{1}-i\frac{1}{1+\lambda^{2}} (K_{1}+\lambda K_{2}),\,\, 
\pr_{2}-i\frac{1}{1+\lambda^{2}}(K_{2}-\lambda K_{1})]=0\;.  \nonumber
\eeqa
Thus
\beqa
K^{\r}_{1}=\frac{1}{1+\lambda^{2}} (K_{1}+\lambda K_{2})\;,\;\;
K^{\r}_{2}=\frac{1}{1+\lambda^{2}}(K_{2}-\lambda K_{1}) \; \nonumber
\eeqa
is a gauge connection with vanishing curvature. Consider 
now an oriented rectangle $R$ constructed by connecting
in order the four 
points $x_{a}=(x_{a}^{1}, x_{a}^{2})=(-L,t)$, $x_{b}
=(L,t)$, $x_{c}=(L,t+ \delta t)$ and
$x_{d}
=(-L,t+\delta t)$. Then the path-ordered
exponential of the line integral of $K^{\r}_{j}$ from
$x_{a}$ to $x_{b}$ to $x_{c}$ to $x_{d}$ and finally back to $x_{a}$
is equal to unity, i.e.
\beq
{\cal P} \exp -i\int_{R} dz \cdot K^{\r}(z) =1\;. \label{3.64}
\eeq
By the boundary condition on $G$ and (\ref{3.64}), the path-ordered
exponential of the line integral of $K^{\r}_{1}$ on the rectangular path
from 
$x_{a}
=(L,t)$ to $x_{b}
=(-L,t)$ must be 
independent of $t$
as $L \rightarrow \infty$. Thus
the expression
\beq
Q(\lambda)={\cal P} \exp -i\int_{-\infty}^{\infty}
dx^{1} \cdot \frac{1}{1+\lambda^{2}} (K_{1}+\lambda K_{2}) 
\eeq 
is a constant 
(that is, independent of $t=x^{2}$) for all $\lambda$. The coefficients
of the Taylor expansion of $Q(\lambda)$
in $\lambda$ are the conserved charges.

\section{Affine connections and the curvature of ${\MM}_{D}$}

With the metric tensor found in section 6, geometric
quantities describing the manifold, such as the
Riemann tensor can be calculated \cite{singer, babelon2}.

To 
make the discussion simpler, the
curvature will first be considered away from points
of ${\MM}_{D}$ where ${\cal D}^{2}$ has zero modes. The
principle value prescription can
then be ignored in (\ref{4.4}) and 
\beqa
G_{(x, j, a) \;(y, k, b)}=\delta_{j\;k} \delta_{a\;b} \delta^{D}(x-y)
-({\cal D}_{j}\frac{1}{{\cal D}^{2}} \;{\cal D}_{k})_{a\;b}\;
\delta^{D}(x-y)\;. \label{simple}
\eeqa 
The
results then agree with references 
\cite{singer, babelon2}.

The derivative of the metric tensor (\ref{simple}) with respect to 
a coordinate is
\beqa
\frac{\delta}{\delta\;A^{l}_{c}(z)}G_{(x, j, a) \;(y, k, b)}
&=&G_{(x, j, a) \;(z, l, d)}(f_{c} \frac{1}{{\cal D}^{2}_{y}}\,
   {\cal D}_{y\;k})_{d\;b} \delta^{D}(x-y)      \nonumber\\
&-&({\cal D}_{x\;j}\,\frac{1}{{\cal D}^{2}_{x}} f_{c})_{a\;d}
\delta^{D}(x-z)\;G_{(z, j, d) \;(y, k, b)}\;, \nonumber
\eeqa
where $(f_{c})ab \equiv f^{acb}$.

Some simple identities (all of which can be proved by
integrating over a distribution and using integration 
by parts) which
are used implicitly in the remainder of this section are
\beqa
{\DD}_{x,j} \frac{1}{ {\DD}^{2}_{x} }
{\DD}_{x,k}\delta^{D}(x-y)
={\DD}_{y,j} \frac{1}{{\DD}^{2}_{y} }
{\DD}_{y,k}\delta^{D}(x-y) \;,        \nonumber
\eeqa
\beqa
\frac{1}{{\DD}^{2}_{x}}
{\DD}_{x,k}\delta^{D}(x-y)
=-\frac{1}{{\DD}^{2}_{y}}
{\DD}_{y,k}\delta^{D}(x-y)    \;,              \nonumber
\eeqa
\beqa
{\DD}_{x,j} \frac{1}{{\DD}^{2}_{x} }
\delta^{D}(x-y)
=-{\DD}_{y,j} \frac{1}{{\DD}^{2}_{y} }
\delta^{D}(x-y) \;,        \nonumber
\eeqa
and
\beqa
G_{(x, j, a) \;(y, k, b)}
=G_{(y, j, a) \;(x, k, b)}   \;.       \nonumber
\eeqa

The triple ``indices" such as $(x, j, a)$, $(y, k, b)$, $(z,l,c)$
and $(u,m,g)$ will sometimes (though not always) be written
as $X$, $Y$, $Z$ and $U$ respectively, and such capital
Roman letters will be used generally to refer to 
a triple consisting of a point of $M$, a coordinate index and
a group index. For example, the notation for a tensor 
(or other) component
$S^{X}_{Y\;Z}$
means $S^{(x, j, a)}_{(y, k, b)\;(z,l,c)}$. The
Einstein summation convention will be used for capital Roman
letters, including integration over $M$, as well as for
the discrete coordinate and group indices. However integration over
$M$ will not be assumed 
when coordinates $x$, $y$, $z$, etc. appear more
than once in an expression. When using capital
Roman indices, functional derivatives will
be written as partial derivatives, e.g.
$\frac{\delta}{\delta A^{a}_{j}(x)}=\partial_{X}$. The components
of the inverse
metric tensor $(G^{-1})^{U\;V}$
(which, as shown earlier is equal to $G_{U\;V}$) will
be denoted $G^{U\;V}$, and $(G^{-1})^{U\;Z}G_{Z\;V}$ will be
written as $G_{U}^{V}$.

Riemannian manifolds with singular metric tensors are discussed
in the appendix. It is shown there that a multi-index object is
a tensor only if it is unaffected by projecting on any of its
indices, e.g.
\beqa
C^{X\;Y\;S}_{Z} G^{Z}_{Q}= C^{X\;Y\;S}_{Q}\;, \;\;
C^{Z\;Y\;S}_{Q} G^{X}_{Z}= C^{X\;Y\;S}_{Q}\;, \nonumber
\eeqa
etc. The covariant derivative must also have this property for
each of its indices.

From the discussion
in the appendix, it follows that the covariant functional
derivative on tensors with upper indices is
\beqa
\nabla _{X} A^{U_{1}\;...\;U_{M}}
&=& \partial_{X} A^{U_{1}\;...\;U_{M}} \nonumber \\
&+& {\Omega}^{U_{1}}_{X V_{1}} \;
    {G^{U_{2}}}_{V_{2}} \cdot \cdot \cdot {G^{U_{M}}}_{V_{M}} 
    A^{V_{1}\;...\;V_{M}} + \cdot \cdot \cdot
       \nonumber \\
&+& {G^{U_{1}}}_{V_{1}} \cdot \cdot \cdot {G^{U_{M-1}}}_{V_{M-1}} 
    {\Omega}^{U_{M}}_{X V_{M}}  \;A^{V_{1}\;...\;V_{M}}  \;. \nonumber
\eeqa
and that on tensors with lower indices is
\beqa
\nabla _{X} A_{U_{1}\;...\;U_{M}}
&=&  \partial_{X} A_{U_{1}\;...\;U_{M}} \nonumber \\
&-&  {\G}^{Z_{1}}_{U_{1} X}\;
     {G^{Z_{2}}}_{U_{2}} \cdot \cdot \cdot 
     {G^{Z_{M}}}_{U_{M}} A_{Z_{1}\;...\;Z_{M}} - 
     \cdot \cdot \cdot  \nonumber \\
&-&  {G^{Z_{1}}}_{U_{1}} \cdot \cdot \cdot 
     {g^{Z_{M-1}}}_{U_{M-1}} 
     {\G}^{Z_{M}}_{U_{M} X}\; 
     A_{Z_{1}\;...\;Z_{M}}    \;,         \nonumber
\eeqa    
where there are two types
of affine 
connections, $\Omega$ and $\G$. The latter
have the
form of Christoffel symbols,
\beqa
{\G}^{U}_{X\;Y}=G^{U\;Z} {\G}_{Z\;X\;Y} \;,\nonumber
\eeqa
where
\beqa
{\G}_{Z\;X\;Y}&=&\frac{1}{2}(\partial_{X} G_{Z\;Y}+\partial_{Y} G_{Z\;X}-
\partial_{Z} G_{X\;Y})  \nonumber \\
&=&\frac{1}{2}[\frac{\delta}{\delta A^{a}_{j}(x)} G_{(z, l, c) \;(y, k, b)}
+\frac{\delta}{\delta A^{b}_{k}(y)} G_{(z, l, c) \;(x, j, a)}
-\frac{\delta}{\delta A^{c}_{l}(z)} G_{(x, j, a) \;(y, k, b)}] \;, \label{d.12}
\eeqa
while the former are
given by
\beqa
{\Omega}^{U}_{V X} 
=-\partial_{X} {G^{U}}_{V} + {\G}^{U}_{V X} \;. \nonumber
\eeqa

The functional derivative of the metric tensor with respect to 
a coordinate is
\beqa
\frac{\delta}{\delta A^{l}_{c}(z)}G_{(x, j, a) \;(y, k, b)}
&=&G_{(x, j, a) \;(z, l, d)}(f_{c} \;\frac{1}{{\cal D}^{2}_{y}}\,
    {\cal D}_{y\;k})_{d\;b} \delta^{D}(x-y) \nonumber 
\eeqa
\beqa
&-&({\cal D}_{x\;j}\;\frac{1}{{\cal D}^{2}_{x}} f_{c})_{a\;d}
    \delta^{D}(x-z)\;G_{(z, j, d) \;(y, k, b)}  \;.     \nonumber 
\eeqa
The reader should keep in mind that these expressions
are not valid on all points of ${\MM}_{D}$; however, they
are valid on the dense set of
points for which the spectrum of $\frac{1}{{\cal D}^{2}}$ is
well-defined. Evaluating (\ref{d.12}) 
yields        
\beqa
{\G}_{Z\;X\;Y}&=& \frac{1}{2} G_{(z,l,c)(x,j,f)} {f}^{afd}
                    (\frac{1}{{\DD}^{2}_{z}}{\DD}_{z,k})_{db}
                    \delta^{D}(z-y)     
                    \nonumber \\                    
            &+& \frac{1}{2} {f}^{afd}
                    ({\DD}_{z,l} \frac{1}{{\DD}^{2}_{z}})_{cd}
                    \delta^{D}(x-z) G_{(x,j,e)(y,k,b)} \nonumber \\ 
            &+& \frac{1}{2} G_{(z,l,c)(y,k,f)} {f}^{bfd}
                    (\frac{1}{{\DD}^{2}_{z}}{\DD}_{z,j})_{da}
                    \delta^{D}(z-x)          \nonumber \\                    
            &+& \frac{1}{2} {f}^{cfd}({\DD}_{z,l}
                    \frac{1}{{\DD}^{2}_{z}})_{cd}
                    \delta^{D}(y-z) G_{(y,k,e)(x,j,a)}   \nonumber \\  
            &-& \frac{1}{2} G_{(x,j,a)(z,l,f)} {f}^{cfd}
                    (\frac{1}{{\DD}^{2}_{x}}{\DD}_{x,j})_{db}
                    \delta^{D}(x-y)             \nonumber \\                    
            &-& \frac{1}{2} {f}^{cde}
                    ({\DD}_{z,l}\frac{1}{{\DD}^{2}_{z}})_{ad}
                    \delta^{D}(x-z) G_{(z,l,e)(y,k,b)} \;. \label{d.13}
\eeqa                                     
Notice that no derivatives act on components of the metric
tensor in any of the terms in this expression. Upon contraction
with $G^{U\;Z}$ to get
${\G}^{U}_{X\;Y}$, the second and fourth terms 
from (\ref{d.13}) vanish. The result
is 
\beqa
{\G}^{U}_{X\;Y} 
&=& \frac{1}{2} \int d^{D} z \; G^{(u,m,g)\;(z,l,c)}
    G_{(z,l,c)\;(x,j,d)} (f_{a} \; \frac{1}{{\cal D}^{2}_{y}}
    {\cal D}_{y,k} )_{d\;b} \delta^{D}(z-y)       \nonumber \\
&+& \frac{1}{2} G^{(u,m,g)}_{(y,k,d)}  
    (f_{b} \; \frac{1}{{\cal D}^{2}_{x}}
    {\cal D}_{x,k} )_{d\;c} \delta^{D}(x-y)       \nonumber \\
&-& \frac{1}{2} \int d^{D} z \; G^{(u,m,g)\;(z,l,c)}
    G_{(z,l,d)\;(x,j,a)} (f_{c} \; \frac{1}{{\cal D}^{2}_{y}}
    {\cal D}_{y,k} )_{d\;b} \delta^{D}(x-y)       \nonumber \\ 
&+& \frac{1}{2} \int d^{D} z ({\cal D}_{x,j}  
    \; \frac{1}{{\cal D}^{2}_{x}} \;f_{c})_{a\;d} \delta^{D}(x-z)
    \;G_{(z,j,d)\;(y,k,b)} G^{(u,m,g)\;(z,j,c)}   \;. \label{d.14}
\eeqa                                     
Notice that
%One can at least partially
%check that (\ref{d.14})
%is the right answer by evaluating the contraction of the indices
%$U$ and $X$. Since ${\sqrt{det^{\prime} G}}=1$,
\beqa
{\G}^{X}_{X\;Y}= 0\;, \nonumber
\eeqa
because contracting the indices in (\ref{d.14}) annihilates the
second and fourth terms, while the first and third terms cancel. 
Furthermore, it is
straightforward to show that $\partial_X G^{Y}_{Y}=0$ and thus
\beqa
{\Omega}^{Y}_{Y\;X}=0\;. \nonumber
\eeqa

As
the reader can check from (\ref{d.14}), contracting both of
the lower indices
of ${\G}$ or $\Omega$ with the inverse metric gives zero, i.e. 
\beqa
{\G}^{X}_{U\;Y} G^{U\;Z}G^{Y\;V}=0 \;, \;\;
{\Omega}^{X}_{U\;Y} G^{U\;Z}G^{Y\;V}=0.  \label{dd.1}
\eeqa                                                            
Furthermore, the partial derivative of $G_{X\;Y}$ also vanishes
when projected thus:
\beqa
G^{X}_{Z} G^{Y}_{V} \partial_{U} G_{X\;Y}=0 \;.       \label{dd.2}
\eeqa

While the proof for manifolds
with singular metric tensors is not as simple
as the nonsingular case, it is true that the covariant 
derivative of the metric tensor is zero. This
is shown in the appendix.

Some understanding of the details
of the spectrum of the 
kinetic term can be obtained
by examining the Ricci curvature \cite{davies}. This is defined
in the usual way as a contraction of the Riemann
tensor. This Riemann tensor was first presented, though
not explained for ${\MM}_{D}$ by
Singer\cite{singer}, and was discussed in some more detail
by Babelon and Viallet \cite{babelon2}. Sometimes performing
tensor
contractions on ${\MM}_{D}$, which requires setting two points
of physical space $M$ equal, requires
ultraviolet regularization. This is true
of the contractions used
in the Laplacian and the Ricci tensor. This point has emphasized by 
Singer \cite{singer}. Singer proposed regularizing
this divergence by inserting a power of the covariant 
Laplacian, $(-{\DD}^{2})^{-s}$, or some other gauge-invariant
damping factor
inside tensor contractions. From
the perspective taken 
here, such a construction, though
probably correct, is less natural than beginning
with                                   
a regularized metric space \cite{kudinov}.

The Riemann curvature tensor
is shown in the appendix
to be
\beqa
R^{U}_{X\;Y\;Z}= {\partial}_{X} {\G}^{U}_{Y\;Z} -
{\partial}_{Y} {\G}^{U}_{X\;Z} + {\Omega}^{U}_{V\;X} {\Omega}^{V}_{Z\;Y}
-{\Omega}^{U}_{V\;Y} {\Omega}^{V}_{Z\;X} \;. \label{d.16}
\eeqa
Notice that for the Yang-Mills case $R_{U\;X\;Y\;Z}=R^{U}_{X\;Y\;Z}$.

The expression (\ref{d.16}) for the Riemann tensor can be simplified
by projecting on the three lower indices and using (\ref{dd.1}) and
(\ref{dd.2}):
\beqa
R^{U}_{X\;Y\;Z}= G^{X^{\r}}_{X} G^{Y^{\r}}_{Y} G^{Z^{\r}}_{Z} 
R^{U}_{X^{\r}\;Y^{\r}\;Z^{\r}}
= G^{X^{\r}}_{X} G^{Y^{\r}}_{Y} G^{Z^{\r}}_{Z} 
({\partial}_{X^{\r}} {\G}^{U}_{Y^{\r}\;Z^{\r}} -
{\partial}_{Y^{\r}} {\G}^{U}_{X^{\r}\;Z^{\r}})  \nonumber \\
= G^{Y^{\r}}_{Y} (\partial_{Y^{\r}} G^{X^{\r}}_{X} G^{Z^{\r}}_{Z} )
{\G}^{U}_{X^{\r}\;Z^{\r}}
-G^{X^{\r}}_{X} (\partial_{X^{\r}} G^{Y^{\r}}_{Y} G^{Z^{\r}}_{Z} )
{\G}^{U}_{Y^{\r}\;Z^{\r}}\; .\label{ddd.2}
\eeqa                                   
Carrying out the differentiations in (\ref{ddd.2})
gives
\beqa
R^{U}_{X\;Y\;Z}= \{ (\partial_{R} G^{S}_{Z}) 
(G^{R}_{Y} G^{T}_{X}-G^{R}_{X} G^{T}_{Y})
+[G^{R}_{Y} (\partial_{R} G^{S}_{X})-
G^{R}_{X} (\partial_{R} G^{S}_{Y}) ] G^{T}_{Z} \}
{\G}^{U}_{T\;S}\; .\label{dd.3}
\eeqa                     
Only the first and third terms in the expression (\ref{d.14}) 
for ${\G}^{U}_{T\;S}$ survive in (\ref{dd.3}). Introducing
the notation
\beqa
B_{s,p}=\frac{i}{{\cal D}^{2}_{s}} {\cal D}_{s,p} \;,
\;\;   {B^{\dag}}_{s,p}= {\cal D}_{s,p} 
\frac{i}{{\cal D}^{2}_{s}}            \;, \nonumber
\eeqa
the Riemann tensor is given, after some more work, by
\beqa
R_{(u,m,g)\;(x,j,a)\;(y,k,b)\;(z,l,c)}= \frac{1}{2} \int d^{D} r\,
\int d^{D} s\, \int d^{D} t\, \int d^{D} v
\nonumber
\eeqa
\beqa
\times 
[G_{(t,q,e^{\r}) (v, i , a^{\r})} \; (f_{b^{\r}}
B_{s,p} )_{a^{\r}\; c^{\r}} \delta^{D}(t-s)    
-G_{(v,i,b^{\r}) (t, q , a^{\r})} \;(f_{f^{\prime}}
\;B_{s,p} )_{a^{\r}\; c^{\r}} \delta^{D}(v-s)] \nonumber 
\eeqa
\beqa
\times
({B^{\dag}}_{s,p})_{c^{\r}\;d} \delta^{D}(s-r) \;G_{(v,i,b^{\r})\;(u,m,d)}
f_{h})_{d\,e}
               \nonumber       
\eeqa
\beqa
\times [ G_{(s,n,e)(z,l,c)} G^{(r,n,h)}_{(y,k,b)} 
G^{(t,q,e^{\r})}_{(x,j,a)}  
-G_{(s,n,e)(z,l,c)} G^{(r,n,h)}_{(x,k,a)} 
G^{(t,q,e^{\r})}_{(y,k,b)}           \nonumber
\eeqa
\beqa
+G_{(s,n,e)(x,j,a)} G^{(r,n,h)}_{(y,k,b)} 
G^{(t,q,e^{\r})}_{(z,l,c)}           
-G_{(s,n,e)(y,k,b)} G^{(r,n,h)}_{(x,j,a)} 
G^{(t,q,e^{\r})}_{(z,l,c)} ] \;, \label{dd.4}
\eeqa
where the primed indices have been introduced so as not to
run out of symbols. This will be recalculated at points of
${\MM}_{D}$
where ${\cal D}^{2}$ has zero modes in another
paper \cite{kudinov}. An important question is whether the
Riemann tensor diverges at these points.

The problem with contracting the indices of (\ref{dd.4}) to
obtain the Ricci curvature is that this involves taking the
trace of operators containing ${\cal P}\,1/{\cal D}^{2}$. Such
expressions are of course
ultraviolet divergent \cite{singer}. Presumably a good
ultraviolet regularization will render ${\MM}_{D}$ compact
(this is definitly
true of the lattice \cite{kudinov}). If a regularization is
introduced, the eigenvalues of the Ricci curvature can be used
to obtain bounds on the ground-state energy 
and information about the heat kernel of
the strongly-coupled theory (i.e. the kinetic term only
of $H$) \cite{davies}. Specifically, one considers a
nonzero displacement vector $\xi^{X}$, which by definition must
satisfy $G^{X}_{Y} \xi^{Y} \neq 0$. Then
what is relevant is the ratio of
quadratic forms
\beqa
Ric(\xi)=\frac{R_{X\;V}  \;\xi^{Y}  \,\, \xi^{U}}
{ G_{W\;Z}\; \xi^{W} \,\, \xi^{Z}} \;. \nonumber
\eeqa 
In particular, one would like to know if $Ric(\xi)$ has
a minimum and if so, whether the minimum is positive. Singer 
\cite{singer} has argued that with
his regularization the answer is affirmative, and concludes
that
the regularized 
Laplacian on ${\MM}_{D}$ should have a pure-point
spectrum.

\section{The Yang-Mills kinetic term}

At last the Laplace operator
on ${\MM}_{D}$ will be worked out
with the metric tensor
\beqa
G_{(x, j, a) \;(y, k, b)}=\delta_{j\;k} \delta_{a\;b} \delta^{D}(x-y)
-({\cal D}_{j}\; {\cal P}\frac{1}{{\cal D}^{2}} \;{\cal D}_{k})_{a\;b}\;
\delta^{D}(x-y)\;, \nonumber
\eeqa 
valid on all of ${\MM}_{D}$. In the appendix, this 
Laplacian on wave functionals
is shown to be
\beqa
\Delta \Psi[\a] 
&=& - \frac{1}{\sqrt{det^{\r}\;G}} \pr_{Y}
   ( {\sqrt{det^{\r}\;G}} \;G^{Y\,U}\pr_{U} \Psi[\a]) \nonumber \\
&+&(\partial_{Z} G^{Z}_{Y})G^{Y\,U}\pr_{U} \Psi[\a]
-\frac{1}{2}(\partial_{Y} G^{Z}_{Z})
G^{Y\,U}\pr_{U} \Psi[\a] \;.         \label{L1}
\eeqa 
It is obvious that the first term
is (\ref{1.55}). However, there are two
new terms present. 

The last term of (\ref{L1}) contains a derivative
of $G^{Z}_{Z}$. This quantity has to be examined with some
care, since it is the trace of an infinite-dimensional
operator and requires
regularization. In a formal sense
\beqa
G^{Z}_{Z}= {\bf 1}^{Z}_{Z}-({\cal D}\; {\cal P} \frac{1}{{\cal D}^{2}}\;
{\cal D})^{Z}_{Z}   \;. \nonumber
\eeqa
The first term is 
\beqa
{\bf 1}^{Z}_{Z}= (N^{2}-1)\; D\; Vol(M)\;, \nonumber
\eeqa
where $Vol(M)$ means the volume of $M$, and is proportional to 
the dimension of 
the space of connections ${\cal U}$, while
the second term is proportional to the dimension
of the space of gauge transformations. Thus $G^{Z}_{Z}$ is
proportional to the dimension of ${\MM}_{D}$. Roughly
speaking, one expects   
\beqa
G^{Z}_{Z}= (N^{2}-1)\; D\; Vol(M)\; 
\frac{\dim  {\MM}_{D}}{\dim {\cal U}} \label{F2}
\eeqa
If the trace is evaluated
carelessly, the result is (since $\epsilon$ is infinitesmal)
\beqa
G^{Z}_{Z}= {\bf 1}^{Z}_{Z}-\frac{1}{2}
\sum_{\pm} ({\cal D}^{2}\; 
\frac{1}{{\cal D}^{2} \pm i \epsilon} )^{Z}_{Z}   
= i \;\epsilon\; tr({\cal P} \frac{1}{{\cal D}^{2}}) =0 \;,\nonumber 
\eeqa
which 
cannot be correct if the
theory
is regularized. In particular, using
a finite lattice \cite{kudinov}, the ratio of dimensions
in (\ref{F2}) will be some positive number. However, what will
be true in any case is that the right-hand
side of (\ref{F2}) is independent of the
physical configuration in ${\MM}_{D}$. Thus 
\beqa
\partial_{Y} G^{Z}_{Z}=0\;. \nonumber
\eeqa

The other unusual 
term in (\ref{L1}) 
is $G^{Y\,U}(\partial_{Z} G^{Z}_{Y} )\partial_{U}$. The
derivative of the metric tensor with respect to 
a coordinate on any point of ${\MM}_{D}$ is
\beqa
\partial_{Z} G^{X}_{Y}=
\frac{\delta}{\delta A^{l}_{c}(z)}G_{(x, j, a) \;(y, k, b)} \nonumber 
\eeqa
\beqa
=G_{(x, j, a) \;(z, l, d)}(f_{c} \;{\cal P}\frac{1}{{\cal D}^{2}_{y}}\,
    {\cal D}_{y\;k})_{d\;b} \delta^{D}(x-y) 
-({\cal D}_{x\;j}\;{\cal P}\frac{1}{{\cal D}^{2}_{x}} f_{c})_{a\;d}
    \delta^{D}(x-z)\;G_{(z, j, d) \;(y, k, b)}            \nonumber 
\eeqa
\beqa
   -\pi^{2} \{ {\cal D}_{x\;j}\; \delta({\cal D}^{2}_{x})
   [{\cal D}_{x\;l}\;f_{c}\; \delta^{D}(y-z)       
+f_{c}\;\delta^{D}(x-z)\;
{\cal D}_{y\;l}]
\delta({\cal D}^{2}_{y}) {\cal D}_{y\;k} \} \delta^{D}(x-y) \;. \label{dd.5}
\eeqa
The factors of $\pi$ can be understood from
the relation 
\beqa
\partial_{Z}  {\cal P}\frac{1}{{\cal D}^{2}_{y}}
= \frac{1}{2} \sum_{\pm} 
\partial_{Z}  \frac{1}{{\cal D}^{2}_{y} \pm i\epsilon}
=-\frac{1}{2} \sum_{\pm}  \frac{1}{{\cal D}^{2}_{y} \pm i\epsilon}
\;\frac{\delta {\cal D}^{2}_{y}}{\delta A^{l}_{c}(z)}\;
\frac{1}{{\cal D}^{2}_{y} \pm i\epsilon}     \nonumber \\
=-\frac{1}{2} \sum_{\pm} 
[{\cal P} \frac{1}{{\cal D}^{2}_{y}} \mp i\pi \delta({\cal D}^{2}_{y})]
\;\frac{\delta {\cal D}^{2}_{y}}{\delta A^{l}_{c}(z)}\;
[{\cal P} \frac{1}{{\cal D}^{2}_{y}} \mp i\pi \delta({\cal D}^{2}_{y})] 
\;. \nonumber
\eeqa
Now if $\partial_{Z} G^{X}_{Y}$ is multiplied by $G^{Y\;U}$, contracting
on $Y$, only the second term of (\ref{dd.5}) survives. Thus 
\beqa
(\partial_{Z} G^{Z}_{Y})G^{Y\,U}=
-\int d^{D} z \int d^{D} x \; \delta^{D}(x-z)
   ({\cal D}_{x\;j}\;{\cal P}\frac{1}{{\cal D}^{2}_{x}} f_{c})_{a\;d}
    \delta^{D}(x-z)\;G_{(z, j, d)}^{( u, k, b)} \;.      \label{L.2}
\eeqa
This quantity is singular and requires
ultraviolet regularization.

The final result of this
section is that the Yang-Mills Hamiltonian is
\beqa
H= \frac{e^{2}}{2} \Delta
+ \int_{M}\,  d^{D}x\; 
\frac{1}{4e^{2}}  \,tr\,F_{j\,k}(x)^{2}\;, \nonumber
\eeqa  
where
\beqa
\Delta \Psi[\a] 
&=& - \pr_{Y}(G^{Y\,U}\pr_{U} \Psi[\a]) +
(\partial_{Z} G^{Z}_{Y})G^{Y\,U}\pr_{U} \Psi[\a]\;, \label{L.4}
\eeqa
and the second term of (\ref{L.4}) can be written
more explicitly with the use of (\ref{L.2}).

\section{Magnetic topography
of ${\MM}_{D}$; scale transformations }

The metric properties of the manifold ${\MM}_{D}$ of configurations
are relevant to the spectrum of the kinetic term of the $SU(N)$ Yang-Mills
Hamiltonian. In order to understand the spectrum at weak coupling, the
potential or magnetic
energy needs to be considered as well. 

A natural starting point is make a relief map
of magnetic energy on ${\MM}_{D}$. The job of
magnetic
topography is not
easy, because there is no obvious relation between this energy
and the
distance from a pure gauge. Nonetheless, using 
some simple scaling arguments, at
least a partial map can made.

The most interesting result of this section 
is an
unexpected feature
of the potential energy function on ${\MM}_{3}$ (in $3+1$ dimensions). It
is found that the
potential energy changes dramatically in response to 
certain small displacements in configuration 
space from a pure gauge. Furthermore
a configuration whose distance from a pure gauge is arbitrarily large
can have a
potential energy which is arbitrarily small. For the unregularized
$2+1$-dimensional Yang-Mills theory, the potential
energy on a sphere of any given
radius in ${\MM}_{2}$, centered around a pure gauge configuration
takes all possible values.

Suppose that the manifold of physical
space, $M$, is very large. Consider 
a rescaling of the coordinates and the connection $A\in {\cal U}$
by a real
factor $s$:
\beq                    
A_{j}(x) \longrightarrow sA_{j}(sx) \;. \label{5.1}
\eeq
This rescaling will take place for $A^{h}$, as long as $h(x)$ is
redefined by
\beq
h(x) \longrightarrow h(sx) \;. \label{5.2}
\eeq
Under (\ref{5.1}) and (\ref{5.2}), the distance from an
equivalence class of pure gauges, $\a_{0}$, transforms
as
\beq
\rho[\Xi(A), \a_{0}] \longrightarrow s^{\frac{2-D}{2}} 
\rho[\Xi(A), \a_{0}]\;. \label{5.3}
\eeq
This follows from the definitions
(\ref{2.1}) and (\ref{2.2}). Note that upon
rescaling, the minimum of ${\sqrt{I[A,B;h,f]}}$ must remain a
minimum (it does not become a maximum or other extremum)
because $s^{\frac{2-D}{2}} >0$.

Suppose that $A\in {\cal U}$ is a particular configuration
of finite potential energy, namely one for
which the magnetic field $F_{jk}(x)$ is continuous
and differentiable, but 
decays rapidly to zero for $x$ outside some
finite bounded region (such as a sphere). This
region
will be called the {\bf domain} of the 
magnetic field. By changing the
size of the domain and the magnitude
of the magnetic field, the distance from some given
pure gauge can be made arbitrarily small (except when 
regularization effects become important) or large (except when 
volume effects become important).

The potential energy
\beqa
U[\Xi(A)]=\int_{M}\,  d^{D}x\; \frac{1}{4e^{2}}  \,tr\,F_{j\,k}^{2}(x) \;, 
\nonumber
\eeqa
transforms as
\beq                                              
U[\Xi(A)] \longrightarrow s^{4-D} U[\Xi(A)]  \;. \label{5.4}
\eeq                                                      
The transformation properties (\ref{5.3}) and (\ref{5.4}) imply
that for a chosen gauge connection $A$ in (\ref{5.1}) and arbitrary
scale factor $s$
\beq
U \sim \rho^{-2\frac{4-D}{D-2}} \;. \label{5.5}
\eeq

Notice that for $2<D<4$ the exponent in (\ref{5.5}) is negative. Thus
it is possible to have arbitrarily large $U$ for arbitrarily
small $\rho$. Since this is true
for any initial choice of $A\in {\cal U}$, there are an infinite
number of physical configurations close to a pure gauge with arbitrarily
large potential energy. This statement has nothing to do with
the gauge group, and holds for the Abelian theory as well.

If a regulator is introduced, the scaling properties
(\ref{5.3}) and (\ref{5.4}) are violated; but these properties
should still hold as long as the domain has a diameter
greater 
than the size of the regulator, such as a lattice 
spacing, $a$. Furthermore, configurations 
near a pure gauge
can have energy
of order $a$, though no greater. There does not
appear to be any obstactle to using (\ref{5.1}), and
a regularization will be implicit in the remainder of
the discussion.

For Abelian gauge theories, other rescalings besides (\ref{5.1})
and (\ref{5.2}) can be considered; these take            
the form
\beq
A_{j}(x) \longrightarrow s^{\phi}A_{j}(sx) \;, \label{5.6}
\eeq
where ${\phi}$ is any real number. Gauge 
invariance can still be maintained under
(\ref{5.6}). This is not
possible for Yang-Mills theories, for which
only $\phi=1$ is permitted. Under (\ref{5.6}) 
\beqa
\rho[\Xi(A), \a_{0}] \longrightarrow s^{\frac{2\phi-D}{2}} 
\rho[\Xi(A), \a_{0}]\;, \nonumber
\eeqa
and
\beqa                                              
U[\Xi(A)] \longrightarrow s^{2\phi+1-D} U[\Xi(A)]  \;. \label{5.8}
\eeqa                                                      
Consequently, by choosing $\phi$ satisfying
\beq
\frac{D-2}{2} < \phi < \frac{D}{2}
\eeq
it is always possible
to make the potential energy arbitrarily small for small $s$, no matter
what the dimension. The configuration then becomes one for
which the diameter of the domain is very large 
but the field strength is very small; a quantum wave functional $\Phi[\a]$
whose amplitude is largest near this configuration is a 
long-wavelength 
photon. This quantum state must
be orthogonal to the vacuum $\Psi_{0}[\a]$ because at least
one of the two wave functionals is 
zero anywhere on ${\MM}_{3}$. This is true since the
distance in configuration space between
the large-domain/low-energy configuration and the pure gauge
configuration 
becomes infinite as $s$ is taken to zero. Not
only are the two states orthogonal, but the energy
eigenvalue of $\Phi[\a]$ becomes zero in this limit. This is
why 
electrodynamics (at least without magnetic monopoles)
has no mass gap in any dimension.                           
               
Is it possible to repeat the argument of the last paragraph
for a Yang-Mills theory? Remarkably, the answer is yes in
$3+1$ dimensions. Notice
that for the 
inequality (\ref{5.8}) to be
satisfied with $\phi=1$, the only
integer solution for the dimension
of $M$ is $D=3$. This evidently
implies that the excitations
of QCD
include
massless particles!

\begin{figure}[t]
\begin{picture}(100,100)(0,0)
\label{2}
\centerline{\epsfbox{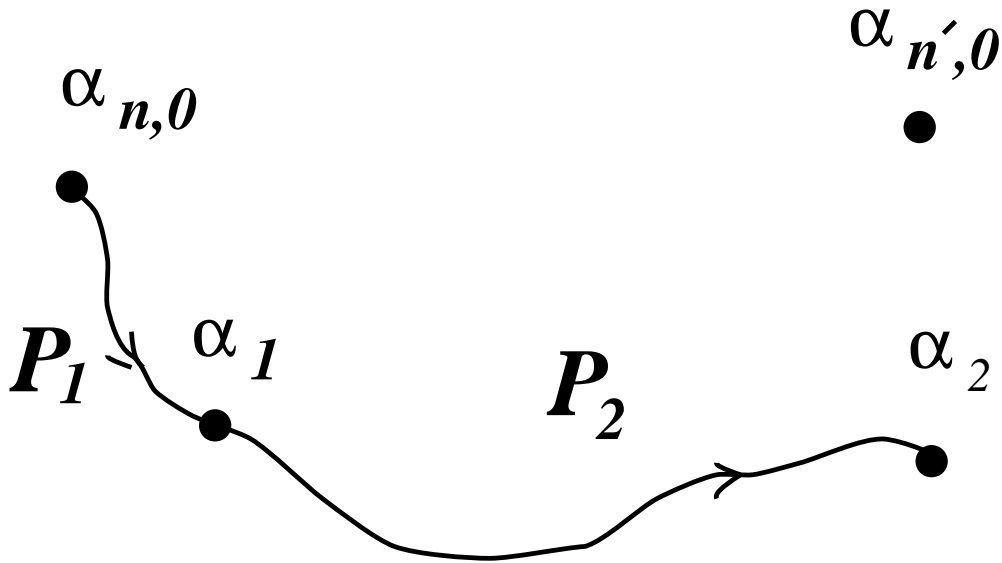}}
\end{picture}
\caption[l]{How to make 
configurations of low potential energy far
from a pure gauge in $3+1$-dimensional Yang-Mills theories. The point 
$\a_{n,0}$ is a pure gauge configuration
of Chern-Simons number $n$. The 
potential energy increases very slightly along the
path $P_{1}$ from $\a_{n,0}$ to $\a_{1}$. The potential
energy decreases along the path $P_{2}$ from
$\a_{1}$ to $\a_{2}$, while
the distance from $\a_{n,0}$
grows.}
\end{figure}

To understand
what is happening in a more detail, consider the following
trajectory of points $\a$ in ${\MM}_{3}$ (figure 2). Start 
from the pure gauge configuration
$\a_{n,0}$ with
a particular Chern-Simons integral $C=n$. Distort the
configuration by increasing the eigenvalues
of magnetic field
strength (which are gauge invariant) to be non-zero
in a small domain. Call this
configuration $\a_{1}$. In doing so, the 
configuration $\a$ moves along
a short path $P_{1}$, directed away from
$\a_{n,0}$, ending at $\a_{1}$, while the
potential
energy $U[\a]$ increases very slightly along this path. Since this
distortion is made 
continuously, $n$ changes only very slightly (it need no longer be an
integer). Now 
perform a scale transformation, with $s$ varying from $1$ to some very
small value. The
potential energy
decreases under this
scale transformation. This has the effect of 
producing a path $P_{2}$ from $\a_{1}$ to a configuration $\a_{2}$
(which is far from $\a_{n,0}$) all the while keeping the potential
energy
arbitrarily small. Furthermore, it
is easy to show by examining (\ref{2.02}) 
that the Chern-Simons integral $C\approx n[\a]$ does
not
change under the scale transformation (\ref{5.1}). If $P_{1}$
and $P_{2}$ are joined to make $P=P_{1}+P_{2}$, the
Chern-Simons integral is nearly unchanged along $P$. The 
possibility of identifying $\a_{2}$ with a
pure gauge other than $\a_{n,0}$ is thereby
ruled out. Thus 
there is no possibility that $\a_{2}$
is a pure gauge.

The wave functional $\Phi[\a]$ which
is large near near $\a_{2}$ is evidently representative
of a long-wavelength
gluon in the quantum theory. It is a state which is
orthogonal to the vacuum and has arbitrarily small energy, by
the arguments used for the Abelian theory. How then can 
there be a gap in the spectrum? Several proposals
are made
in the next section.

In $2+1$ dimensions, the situation is very different. There
is only one point of configuration space ${\MM}_{2}$ which
is an equivalence class of pure gauge fields, namely
$\Xi(0)$. Under a rescaling
(\ref{5.1}), the potential energy of
some configuration $\a$
is proportional to $s^{2}$, but
the distance from $\Xi(0)$ does not change 
at all. The
scale transformation moves
the configuration around
a sphere in ${\MM}_{2}$, centered around $\Xi(0)$. The range
of values of the potential
energy on this sphere are all real 
numbers between zero and infinity.

\section{Magnetic topography
of ${\MM}_{D}$; the mass gap }

The physical implications of section 10 will now
be examined.

Perhaps the first counter-argument against concluding that
there is no gap in $3+1$ dimensions
which comes to
mind is that the
potential energy function is extremely
``hilly", rising and falling rapidly, on all
of ${\MM}_{3}$ for Yang-Mills theories (as it certainly is near
the origin). Then there is the possibility
of states which are localized in ${\MM}_{3}$, as
occurs for a random potential, even though
there are some directions in which the potential energy
falls rapidly. The regions of small potential energy
would have large electric
(kinetic) energy by the uncertainty principle. In
this
way, the first excited state could have a finite gap
above the ground state. Even if
such potential-energy barriers exist, they would all
fade away as a scale transformation (\ref{5.1}) is made. However, the
uncertainty principle could make the kinetic energy increase
under such the scale transformation. Such a
mechanism is very close to the
picture of extrapolating the effective
coupling to large distances 
\cite{magnetic}; the renormalized potential would
then grow under scaling transformations (\ref{5.1}) rather than
shrink. To investigate this
possibility requires exploration of the kinetic energy
(the Laplacian on ${\MM}_{3}$) and the potential energy
in a small neighborhood of $\a_{2}$. It is left as an exercize
for the reader to show that formally, i.e. ignoring 
regularization and renormalization, the Riemann curvature
scalar decreases with decreasing 
$s$, $R[\a] \rightarrow s^{D-2}\;R[\a]$, suggesting that
curvature effects aren't very
important at large scales. Of course, this relation
has to be modified by an anomalous dimension. Thus it 
may be that the quantum theory has a gap to
the first excited state, though simple intuition about
the classical theory would indicate otherwise. Some
examples of quantum-mechanical systems for which this
is the case were discussed by Simon in reference \cite{simon}
(including a model suggested by Goldstone and Jackiw
motivated by the Yang-Mills theory).

Another possible explanation for the gap
is that the true vacuum is
a superposition of states of different Chern-Simons 
integrals \cite{thetavac}. While
there is an energy barrier between $\a_{2}$ and a pure
gauge configuration $\a_{n^{\prime},0}$ of
Chern-Simons integral $n^{\r}\neq n$, the distance between 
them, $\rho[\a_{n,2},\a_{n^{\prime},0}]$
could conceivably be very small. If this is the case, it is
not possible to create a state orthogonal to the ground
state by the arguments above. This is because
the quantum state associated
with $\a_{2}$ would have a significant overlap
with the true vacuum state, which is not an eigenstate
of the Chern-Simons integral. A general
argument
could be made if an upper bound on $\rho[\a_{2},\a_{n^{\prime},0}]$ 
could be found.

%In
%fact, it seems almost obvious that this is the case from
%the following\footnote{I thank Jan Ambj{\o}rn for
%suggesting this line of argument}. Recall
%that the action of a Euclidean instanton from Chern-Simons
%number $n$ to $n^{\r}$ does
%not depend on its size $\lambda$. The action
%of the instanton, $\vert n- \n^{\r} \vert S_{0}$,
%is roughly $T\;\Delta E$, where $T$ is 
%the time of the tunneling transition and $\Delta E$ is the change in
%energy. At large scales (i.e. large $\lambda$) the
%transition time $T \sim \lambda$ grows and $\Delta E$ vanishes. But
%also at sufficiently
%large scales, the 
%potential energy can be neglected (from (\ref{5.4}). This
%means that along
%some path from the
%configuration $\a_{n,2}$ to a configuration with Chern-Simons number
%$n^{\r}$ the kinetic energy is extremely small. 

In my opinion, both of the 
mechanisms invoked above for a gap are  probably
correct for finite
$N$. In the large-$N$ limit \cite{largeN}, the importance of the
latter may be diminished, as tunnelings between
different Chern-Simons numbers are suppressed \cite{witten}.

One might also guess that the volume of regions in ${\MM}_{3}$
far from $\a_{n,0}$ is small, making
such regions 
inaccesible. However, this guess
is false. To see this, consider some set of
points a constant distance from $\a_{n,0}$. Under the
rescaling, the distance between any pair of
points in this set grows at the same rate as the distance
between either point in the pair and $\a_{n,0}$. Thus spheres
in ${\MM}_{3}$ always increase in size under a rescaling. This
is consistent with a decreasing Riemann curvature
scalar for decreasing $s$.

It was shown that for the 
$2+1$-dimensional Yang-Mills theory, the potential
energy on those 
points in ${\MM}_{2}$ which are
a fixed distance from $\Xi(0)$
takes all possible values. Note that when $D<2$, the scaling
formula (\ref{5.5}) indicates
that that the potential
energy
of a configuration {\em increases} with
its distance from $\Xi(0)$ 
(in fact, for $D=2-\epsilon$ the distance decreases as
as $s\rightarrow 0$, which is consistent with a gap). It 
should be true that the
distance function $\rho[\cdot,\cdot]$ can be defined in the 
{\em renormalized}
$2+1$-dimensional Yang-Mills theory (indeed, this
was Feynman's viewpoint \cite{feynman}), and that this property
is present there as well. If so, a mass gap should
be present. Since
the theory is super-renormalizable, it should be much easier
to study the renormalized metric and potential
energy than in $3+1$ dimensions.

What happens when the space dimension, $D$, is greater than $3$? The 
potential
is constant under a rescaling
for ${\MM}_{4}$, suggesting a massless
phase at weak coupling. However for $D=5$, the 
scaling formula again
(\ref{5.5})
suggests (but does not prove! See below) that the energy
of a configuration increases with
its distance from a pure gauge. This suggests
that the $5+1$-dimensional 
gauge theory has a mass gap. This theory
almost
certainly does not confine at weak coupling, which means
that it would exhibit a massive
Higgs phase \cite{thooft} when the bare coupling
constant is between zero and some value. As far as I 
know, lattice Monte-Carlo calculations have not ruled
out this possibility. To be certain that 
the potential energy
function
is convex, it is neccesary to investigate other motions
in ${\MM}_{5}$ besides (\ref{5.1}) to see whether
that the potential could decrease in other directions.

\section{Discussion}

What was done was to
start at the bottom from analysis and work up to geometry. The
result is at least the barest outlines
of the space in which physical Yang-Mills configurations live. Through
a careful study of the
distance function, the
kinetic
energy operator in the Schr{\"o}dinger representation
was found on this space. The
magnetic energy on this space
was investigated for different dimensions of space-time. 

The point of view taken here
was different from that of references
\cite{zwanziger, van baal, cutkosky, fujikawa, fuchs} in
that no gauge fixing (except
for the original temporal gauge condition)
was imposed. This viewpoint
has also been advocated in reference 
\cite{friedberg}. The drawback of
a metric tensor with zero eigenvalues seems outweighed
by the advantage of not having to isolate
a Gribov domain. To make an
analogy with Riemann surfaces, the methods
used here and in reference \cite{friedberg}
are like the use 
of automorphic functions, while what the authors
of references 
\cite{zwanziger, van baal, cutkosky, fujikawa, fuchs}
have been attempting to do is like 
directly determining the properties of
the fundamental region.

It should not be very hard to
introduce similar techniques for Euclidean path
integrals. Then the analysis here is relevant 
to the $D$-dimensional Yang-Mills theory with
no gauge fixing, rather that the $D+1$-dimensional
Hamiltonian theory with temporal gauge.

There
are many issues
to examine
further. The metric, Laplacian and Ricci
curvature need to be carefully recomputed 
with some explicit regularization \cite{kudinov}. The
issue of the measure and reducible connections \cite{fuchs}
should also be rexamined on the lattice. Closed
geodesics and conjugate points will be
very interesting 
to study \cite{gribov, singer, babelon1, babelon2}. The 
distance 
between configurations
$\a_{2}$ and $\a_{n^{\prime},0}$, discussed in section
11, must be estimated. What
is most important is that
displacements in ${\MM}_{D}$ other than those
discussed in sections 10 and 11 should be investigated.

The $3+1$-dimensional low-energy configuration
found in section 10 is not a 
{\em perturbative} gluon. The former is a large geodesic distance
from a pure gauge
in ${\MM}_{3}$, whereas the latter is a small geodesic
distance from a pure
gauge.

The
metric for the
super-renormalized theory in $2+1$ dimensions needs to
be studied. It may be that it is easier than in $3+1$ dimensions
to look at the potential in directions perpendicular to the
scale direction studied in section 10. In fact it may be
even more straightforward to study the theory in just above two
space-time dimensions, i.e. in dimension $(1+\epsilon)+1$ 
\cite{dineetal}, for which the number of degrees of freedom might
be 
small enough to find a complete picture of the Laplacian
and the potential
energy function on all of ${\MM}_{1+\epsilon}$.

\section*{Appendix: Differential calculus on manifolds
with singular metric tensors}

For the reader who is uncomfortable
with the notion of a singular metric
tensor discussed in this paper, the general
finite-dimensional
case is discussed here. An example is given which
describes
the distance function on a two-dimensional sphere. Like
the metric discussed for Yang-Mills fields, it may be used
globally on the manifold; it is not neccesary to
introduce additional coordinate charts.

Consider a manifold for which the dimension of tangent 
space, $d$, is a fixed number, and for which
the number of coordinates, $N$, of any coordinate chart
is also a fixed number. The metric tensor will be singular
provided $d<N$. Label the coodinates by $x=(x^{1},...,x^{N})$
in some given coordinate chart and label
some choice of basis of tangent space at $x$ by 
${\bf e^{1}}(x),...{\bf e^{d}}(x)$. Since
the coordinate chart is $N$-dimensional, there are neccesarily vector
fields which are othogonal to tangent space; thus for some functions
of coordinates $\phi(x)$
\beqa
\nabla \phi(x)   \cdot    {\bf e^{J}}(x)=0,\;J=1,...,d\;\;.  \label{AA1}
\eeqa
In order for (\ref{AA1}) to be consistent, a point $x$ in a coordinate
chart ${\cal C}$ and $x+dx \cdot \nabla \phi(x)$ must be 
identified. This means that points of the
manifold cannot be parametrized by points in ${\cal C}$, but rather
by curves in ${\cal C}$ for $N-d=1$, by surfaces in
${\cal C}$ for $N-d=2$, etc. Experts in geometry will see that these
curves, surfaces, etc. are just fibers of an $N$-dimensional bundle.

The coordinate-space
components of the tangent vector ${\bf e}^{J}(x)$
will be denoted by ${e^{J}}_{j}$. A tangent-space vector field
can be written ${\bf A}(x)$ and has the standard
representation
\beqa
{\bf A}(x)= \sum_{J} {\bf e}^{J}(x) A_{J}(x)   \;. \label{AA2}
\eeqa 
Unfortunately, as it stands, $(\ref{AA2})$
is not very useful, because the components $A_{J}(x)$
are not coordinate components. The latter can be found after
first defining an $N$-dimensional set of vectors
${\bf E_{1}}(x),...{\bf E_{N}}(x)$; these span a vector space
which includes tangent space as a subspace. Then the 
vector field ${\bf A}(x)$ (henceforth called
a vector) can be written
as
\beqa
{\bf A}(x)= \sum_{j} {\bf E_{j}}(x) A^{j}(x)  \;. \label{AA3}
\eeqa 
However, (\ref{AA3}) does not show explicitly that 
${\bf A}$ is a tangent vector. 

For the purposes of both this appendix and the text
of this article it 
will be assumed that 
\beq
{\bf E}_{i}(x) \cdot {\bf E}_{j}(x)=\delta_{ij}\;. \label{AA4}
\eeq
In 
general, one can 
consider cases where (\ref{AA4}) is not so; the only
complication is more index clutter, and there is
little essential difference. Clearly ${e^{J}}_{j}
={\bf E}_{j} \cdot  {\bf e}^{J}(x)$.

The metric tensor is defined by
\beqa
g_{ij}(x)= \sum_{K}{e^{K}}_{i}(x)  
{e^{K}}_{j}(x)  \;. \nonumber
\eeqa
This
is the correct definition because
a change in coordinates $dx$ produces a displacement $dY$ 
in tangent space 
\beqa
dY^{K}=\sum_{i}{e^{K}}_{i}(x)dx^{i} \;.
\eeqa
The arc-length differential
$ds$ given by
\beq
ds^{2}= \sum_{K} dY^{K} dY^{K} =g_{ij}\;dx^{i} dx^{j}
\eeq                        
Clearly
$g_{ij}(x)$ is singular. 

There is a meaningful
notion of  inverse metric
tensor. Consider now another basis
of the $d$-dimensional tangent space, namely the the orthonormal basis 
of eigenvectors of the matrix $g_{ij}(x)$ with nonzero eigenvalues
${\bf S}_{1}(x),...,{\bf S}_{d}(x)$, ${\bf S}_{q}(x) \cdot {\bf S}_{r}(x)
=\delta_{qr}$. The metric
tensor can be written as
\beqa
g_{ij}(x)= \sum_{J}  \lambda_{J} {S^{J}}_{i}(x) {S^{J}}_{j}(x) \;, \label{C1}
\eeqa
where $\lambda_{J} \neq 0$. The
projection operator into tangent space is
\beq
{P^{i}}_{j}(x) =  
\sum_{J} { S^{J}}_{i}(x){S^{J}}_{j}(x)\;. \label{C2}
\eeq
The mismatch of upper and lower indices in (\ref{C2}) is not
a misprint. The expression for the inverse metric tensor
\beqa
g^{ij}=\sum_{J}  (\lambda_{J})^{-1} { S^{J}}_{i}{S^{J}}_{j} \;, \label{C3}
\eeqa
also displays such a mismatch. Notice
that
\beq
{P^{i}}_{j} = g^{ik}g_{kj} \equiv {g^{i}}_{j} \neq \delta^{i}_{j}\;,\label{AA9}
\eeq
where the Einstein summation convention has been used. Important
differences arise between some of the expressions found
here and those of standard Riemannian geometry because
\beq
\partial_{l}   {g^{i}}_{j} \neq 0  \;. \label{AA10}
\eeq

Tensor fields on the manifold are
mappings into the real numbers from outer products
of the tangent space, $T$, and its dual vector
space $T^{*}$. Furthermore, if a displacement is
made along a zero eigenvector
of the metric $g_{ij}$, this mapping must not
change. This means that there are
two neccesary and sufficient
conditions for a general
expression like ${C^{ij}}_{klm}(x)$ to be a tensor field
(henceforth called a tensor). These are: \begin{itemize}
\item {\bf condition 1.} a tensor is not changed by contracting any
index with that of the projection, e.g.
\beqa
{C^{ij}}_{klm} = {g^{n}}_{k}  {C^{st}}_{nrq}  \;, \nonumber
\eeqa                         
and
\beqa
{C^{ij}}_{klm} = {g^{i}}_{s}  {C^{st}}_{nrq}  \;. \nonumber
\eeqa                         
\item {\bf condition 2.} defining the object (which lies
in the space $T \otimes T \otimes T^{*} \otimes T^{*} \otimes T^{*}$)
\beqa
{\bf C} \equiv {C^{ij}}_{klm} g^{kr} g^{ls} g^{m t}
{\bf e}_{i} \otimes {\bf e}_{j}  \otimes {\bf e}_{r}  \otimes 
{\bf e}_{s}  \otimes {\bf e}_{t} \;, \nonumber
\eeqa
its derivative must satisfy
\beqa
({\delta^{l}}_{m}-{g^{l}}_{m} )\partial _{l} {\bf C}=0\;. \nonumber
\eeqa
\end{itemize}

The metric tensor obviously satisfies condition 1. Consistency
demands that any example must also satify condition
2. If this is the case, the covariant derivative 
of the metric tensor is zero, as will be shown below.

In Riemannian geometry, the
covariant derivative of a tensor
is the coordinate
derivative of that
tensor projected into tangent space. The
definition of the covariant derivative on tensors with only upper
indices (that is tensors which act on outer
products
of tangent space) is
\beqa
\nabla _{l} A^{p_{1}\;...\;p_{M}}
&=&      g^{p_{1}k_{1}} \cdot \cdot \cdot  g^{p_{M}k_{M}}
         \sum_{K_{1}...K_{M}} {e^{K_{1}}}_{k_{1}} \cdot \cdot \cdot 
        {e^{K_{M}}}_{k_{M}} \nonumber \\
&\times& \partial_{l} ({e^{K_{1}}}_{m_{1}} 
\cdot \cdot \cdot {e^{K_{M}}}_{m_{M}} 
A^{m_{1}\;...\;m_{M}}) \;. \label{AA12}
\eeqa
For the case of
a singular metric, $A_{p_{1}\;,...\;p_{M}}$ is not a tensor
unless condition 2,
\beqa
\nabla_{j} A^{p_{1}\;...\;p_{M}} = {g^{l}}_{j}  
\nabla _{l} A^{p_{1}\;...\;p_{M}} \;, \nonumber
\eeqa
is satisfied. Carrying 
out the differentiations in (\ref{AA12}) yields
\beqa
\nabla _{l} A^{p_{1}\;...\;p_{M}}
&=& {g^{p_{1}}}_{m_{1}} \cdot \cdot \cdot {g^{p_{M}}}_{m_{M}} 
    \partial_{l} A^{m_{1}\;...\;m_{M}} \nonumber    \\
&+& {\G}^{p_{1}}_{l m_{1}} \;
    {g^{p_{2}}}_{m_{2}} \cdot \cdot \cdot {g^{p_{M}}}_{m_{M}} 
    A^{m_{1}\;...\;m_{M}} + \cdot \cdot \cdot
       \nonumber \\
&+& {g^{p_{1}}}_{m_{1}} \cdot \cdot \cdot {g^{p_{M-1}}}_{m_{M-1}} 
    {\G}^{p_{M}}_{l m_{M}}  \;A^{m_{1}\;...\;m_{M}}  \;, \label{AA14}
\eeqa
where
\beqa
{\G}^{p}_{lm} 
=g^{p k} [\sum_{K} {e^{K}}_{k} 
   \partial_{l} {e^{K}}_{m}] \;. \label{AA15}
\eeqa  
By the usual reasoning, (\ref{AA15}) implies
Christoffel's formula 
\beqa
{\G}^{p}_{lm} = \frac{1}{2} g^{p k}
(\partial_{l} g_{km} 
+\partial_{m} g_{kl} 
-\partial_{k} g_{lm}) \;.\nonumber
\eeqa

The reader should notice that the covariant derivative above
resembles
the standard covariant derivative. However, because of (\ref{AA10}), this
is not true of the covariant
derivative on a tensor
with lower indices, $\nabla _{l} A_{k_{1}\;...\;k_{M}}$. This is
\beqa
\nabla _{l} A_{k_{1}\;...\;k_{M}}
&=&        \sum_{K_{1}...K_{M}} {e^{K_{1}}}_{k_{1}} \cdot \cdot \cdot 
        {e^{K_{M}}}_{k_{M}} 
          \partial_{l} ({e^{K_{1}}}_{m_{1}} 
         \cdot \cdot \cdot {e^{K_{M}}}_{m_{M}} \nonumber \\
&\times& g^{m_{1}r_{1}} \cdot \cdot \cdot 
g^{m_{M}r_{M}} A_{r_{1}\;...\;r_{M}}) \;. \nonumber
\eeqa
This may be written as
\beqa
\nabla _{l} A_{k_{1}\;...\;k_{M}}
&=&   {g^{r_{1}}}_{k_{1}} \cdot \cdot \cdot {g^{r_{M}}}_{k_{M}}
      \partial_{l}  A_{r_{1}\;...\;r_{M}} \nonumber \\
&-&  {\Omega}^{r_{1}}_{k_{1} l}\;
     {g^{r_{2}}}_{k_{2}} \cdot \cdot \cdot 
     {g^{r_{M}}}_{k_{M}} A_{r_{1}\;...\;r_{M}} - 
     \cdot \cdot \cdot  \nonumber \\
&-&  {g^{r_{1}}}_{k_{1}} \cdot \cdot \cdot 
     {g^{r_{M-1}}}_{k_{M-1}} 
     {\Omega}^{r_{M}}_{k_{M} l}\; 
     A_{r_{1}\;...\;r_{M}}    \;,         \label{AA18}
\eeqa    
where the affine
connection now takes a different form:
\beqa
{\Omega}^{r}_{k l} =-\sum_{K} {e^{K}_{k}} \partial_{l} ({e^{K}}_{m} g^{mr})
=-\partial_{l} {g^{r}}_{k} + {\G}^{r}_{k l} \;, \label{AA19}
\eeqa 
which is not symmetric in the indices $k$ and $l$. If 
$A_{r_{1}\;...\;r_{M}}$ is a tensor, by condition 2,
\beqa
\nabla_{j} A_{k_{1}\;...\;k_{M}} = {g^{l}}_{j}  
\nabla _{l} A_{k_{1}\;...\;k_{M}} \;. \nonumber
\eeqa

The covariant
derivative of tensors with
mixed upper and lower indices can be defined similarly. At any
rate, indices of tensors can be lowered and raised by the metric
tensor and its inverse, respectively.

There are more convenient expressions for covariant derivatives than
those given above. Pushing 
the products of projections ${g^{p_{1}}}_{m_{1}} \cdot 
\cdot \cdot {g^{p_{M}}}_{m_{M}}$ past the partial derivative
in (\ref{AA14}) gives
\beqa
\nabla _{l} A^{p_{1}\;...\;p_{M}}
&=&  \partial_{l} A^{p_{1}\;...\;p_{M}} \nonumber    \\
&+& {\Omega}^{p_{1}}_{m_{1} l } \;
    {g^{p_{2}}}_{m_{2}} \cdot \cdot \cdot {g^{p_{M}}}_{m_{M}} 
    A^{m_{1}\;...\;m_{M}} + \cdot \cdot \cdot
       \nonumber \\
&+& {g^{p_{1}}}_{m_{1}} \cdot \cdot \cdot {g^{p_{M-1}}}_{m_{M-1}} 
    {\Omega}^{p_{M}}_{m_{M} l}  \;A^{m_{1}\;...\;m_{M}}  \;. \label{AA21}
\eeqa
Similiarly (\ref{AA18}) can be rewritten as
\beqa
\nabla _{l} A_{k_{1}\;...\;k_{M}}
&=&  \partial_{l} A_{k_{1}\;...\;k_{M}} \nonumber \\
&-&  {\G}^{r_{1}}_{k_{1} l}\;
     {g^{r_{2}}}_{k_{2}} \cdot \cdot \cdot 
     {g^{r_{M}}}_{k_{M}} A_{r_{1}\;...\;r_{M}} - 
     \cdot \cdot \cdot  \nonumber \\
&-&  {g^{r_{1}}}_{k_{1}} \cdot \cdot \cdot 
     {g^{r_{M-1}}}_{k_{M-1}} 
     {\G}^{r_{M}}_{k_{M} l}\; 
     A_{r_{1}\;...\;r_{M}}    \;,         \label{AA22}
\eeqa    

If the metric tensor
is truly a tensor, meaning that it satisfies conditions
1 and 2 above, its covariant derivative 
vanishes. To prove this for singular
metric tensors is a little more tedious than in the
nonsingular case. Now
\beqa
\nabla_{l} g_{k\;n}
&=&\partial_{l} g_{k\;n}
   -{\G}^{r}_{k\;l} g_{r\;n}-{\G}^{s}_{n\;l} g_{k\;s} \nonumber \\
&=& [\partial_{l} g_{k\;n} -\frac{1}{2} g^{s}_{k} \partial_{l} g_{s\;n}
    -\frac{1}{2} g^{s}_{n} \partial_{l} g_{s\;k}] \nonumber \\
&-& \frac{1}{2}[ g^{s}_{r} \partial_{k} g_{s\;l} 
      -g^{s}_{k} \partial_{s} g_{r\;l}]           \nonumber \\
&+& \frac{1}{2}[ g^{s}_{r} \partial_{s} g_{k\;l} 
      -g^{s}_{k} \partial_{r} g_{s\;l}]    \;.       \label{AAAA}
\eeqa
Contracting this derivative 
with projection matrices must leave it unchanged:
\beqa
\nabla_{l} g_{k\;n} = 
g^{a}_{k} g^{b}_{n} \nabla_{l} g_{a\;b}\;. \label{BBBB}
\eeqa
Applying (\ref{BBBB}) to each of the terms in brackets $[\;]$
in (\ref{AAAA}) gives zero.

The first guess for the volume element is 
${\sqrt{\;det^{\r} \;g}}$, where the
prime means that the zero
eigenvalue has been removed from the determinant. However, this 
cannot always be correct, because
the integration $\int d^{N} x$ is over the entire coordinate space and
not the $d$-dimensional manifold. In 
situations where the $N-d$ integration in the singular
directions has a volume which is independent of the position
on the manifold, this produces
a (possibly infinite) constant factor, which can be 
divided out.  For the example given in this appendix below and for
the Yang-Mills theory, this turns out to be the case.

Consider a real-valued field $\phi(x)$. The derivative
(which is the covariant derivative for a 
scalar), $\partial_{k} \phi$, 
and the projection of the 
derivative, ${g^{l}}_{k}\partial_{l} \phi$ must
be the same, by condition 2 for the $0$-tensor $\phi$. The
Laplacian is
\beqa
\Delta \phi = -{\bf \nabla}_{j} g^{jk}\partial_{k} \phi=
-(\pr_{j}+{\Omega}^{k}_{j\,k})g^{j\,l}\pr_{l} \phi 
\;. \nonumber
\eeqa
The Laplacian can be brought into a more convenient form
using (\ref{C1}), (\ref{C2}) and (\ref{C3}). Using these 
expressions, the contraction of the connection ${\G}^{k}_{j\,k}$ 
may be written
as
\beqa
{\G}^{k}_{j\,k}= \sum_{J, \lambda_{J} \neq 0} 
(\frac{1}{2}\frac{\partial_{j} \lambda_{J}}{\lambda_{J} }+\sum_{k}
S^{J}_{k} \partial_{j} S^{J}_{k} )
=\frac{1}{\sqrt{det^{\r}\;g}} \partial_{j}  {\sqrt{det^{\r}\;g}}
+\frac{1}{2} \partial_{j} g^{k}_{k}\;. \nonumber
\eeqa 
Therefore
\beqa
{\Omega}^{k}_{j\,k}=\frac{1}{\sqrt{det^{\r}\;g}} 
\partial_{j}  {\sqrt{det^{\r}\;g}}
+\frac{1}{2}\partial_{j} g^{k}_{k}-\partial_{k} g^{k}_{j}\;. \nonumber
\eeqa 
The Laplacian is consequently
\beqa
\Delta \phi =
-\frac{1}{\sqrt{det^{\r}\;g}} \pr_{j}
({\sqrt{det^{\r}\;g}} g^{j\,l}\pr_{l} \phi)
+(\partial_{k} g^{k}_{j}-\frac{1}{2}\partial_{j} g^{k}_{k})
g^{j\,l}\pr_{l} \phi 
\;.                                                       \label{A10}
\eeqa                                   
This is identical to the expression for the Laplacian
in the nonsingular
case, except for an additional term which depends on the
derivatives of the projection matrix $g^{k}_{j}$.

The curvature is defined as the commutator of two covariant
differentiations on a vector
$\xi^{l}$. The definition does not need to be 
modified, because covariant differentiations of tensors are
infinitesmal displacements in nonsingular directions (by
condition 2 above). The Riemann curvature tensor is
\beqa
R^{k}_{ijl} \xi^{l} =[\nabla_{i} , \nabla_{j}] \xi^{k} \;. \label{AB1}
\eeqa
applying (\ref{AA21}) and (\ref{AA22}) to (\ref{AB1}) gives
\beqa
R^{k}_{ijl}
&=& \partial_{i} \Omega^{k}_{lj}- \partial_{j} \Omega^{k}_{li}
    +\Omega^{k}_{si} \Omega^{s}_{lj}
    -\Omega^{k}_{sj} \Omega^{s}_{li} \nonumber \\
&=& \partial_{i} \G^{k}_{lj}- \partial_{j} \G^{k}_{li}
    +\Omega^{k}_{si} \Omega^{s}_{lj}
    -\Omega^{k}_{sj} \Omega^{s}_{li} \;, \nonumber
\eeqa
where (\ref{AA19}) was used in the last step. As usual, the
the Ricci tensor is $R_{il}=R^{k}_{ikl}$, and the curvature
scalar is $R=g^{il}R_{il}$.

As an application of these concepts, consider
the two-dimensional unit sphere, with
coordinates $x^{1}, x^{2}, x^{3}$, satisfying
$-\infty < x^{j} <\infty$, excluding the point
${\bf x}=0$. Then ${\bf {\hat x}}={\bf x}/\vert {\bf x} \vert$ is
a point
on this sphere. The metric
is the length of
a chord connecting two points
on the sphere. Thus if two three-vectors are ${\bf x}$ and
${\bf y}$, the distance between them, $s({\bf x}, {\bf y})$, is
given by
\beqa
s({\bf x}, {\bf y})^{2}=(\frac{{\bf x}}{\vert {\bf x} \vert}
-\frac{{\bf y}}{\vert {\bf y} \vert} )^{2}\;. \nonumber
\eeqa
Thus 
\beqa
ds^{2}=s({\bf x}+d{\bf x}, {\bf x})^{2}=g_{ij} dx^{i} dx^{j} \nonumber
\eeqa
where 
the metric tensor 
is the three-by-three 
matrix
\beq
g_{i j}= \frac{1}{{\bf x} \cdot {\bf x}} (\delta_{i j}-
\frac{x^{i} x^{j}}{{\bf x} \cdot {\bf x}}) . \label{A1}
\eeq
While lower indices are used on the (doubly-covariant)
metric
tensor, raised indices are used on the coordinates
on the right-hand-side of (\ref{A1}). Since this
equation is just an expression
for the
components of the metric tensor, this
should cause no confusion. The reader can easily verify
that the matrix $g$ is singular. Three orthogonal
eigenvectors
are ${\bf x}$, any vector ${\bf y}$ which is perpendicular
to ${\bf x}$ and ${\bf z}={\bf x} \times {\bf y}$. The eigenvalues
of ${\bf x}$, ${\bf y}$ and ${\bf z}$ 
are $0$, $({\bf x} \cdot {\bf x})^{-1}$ and 
$({\bf x} \cdot {\bf x})^{-1}$, respectively. 

Notice that the manifold is {\em not} the
two-dimensional real-projective
space $RP_{2}$. In $RP_{2}$ the point
${\bf x}\in R^{3}$ is identified with $\lambda {\bf x}$ for
any real $\lambda \neq 0$. The surface considered here
is different in that $\lambda$ is actually positive. The
antipodes of the sphere are not identified.

It is straightforward to check
that (\ref{A1}) gives the metric of 
the two-dimensional sphere after
the substitution of polar coordinates
$r$, $\theta$ and $\phi$ through
\beqa
{\bf x}=r(\sin\theta \cos \phi, \sin \theta \sin \phi, \cos \theta) \;.
\nonumber 
\eeqa
The infinitesmal distance is then
\beqa
ds^{2}=d\theta^{2}-\sin^{2} \theta d\phi^{2}\;.
\eeqa
The geometric reason for the singularity of $g$ is clear; the
distance function does not depend on 
${\sqrt {{\bf x} \cdot {\bf x}}}=r$. The
meaning of this is that all the concentric
spheres around the origin have been identified with one another. 

At this stage, the reader might think that there is
really no distinction between the singular metric formulation
of the sphere and the standard embedding method of introducing
a mapping 
\beq
\theta, \phi \longrightarrow x^{1}, x^{2}, x^{3}\;,
\label{A100}
\eeq
and defining the intrinsic geometric quantities in terms
of the derivatives of this mapping. The 
new feature
is that 
two-dimensional coordinate charts
are never introduced. In
the usual approach, the embedding (\ref{A100}) is used
to construct the normal bundle, which is a fiber bundle for the case of
a sphere $S_{2}$ in a double covering
of $R^{3}$ (as well as for the case
of the manifold of gauge configurations). The singular metric
technique makes it unneccesary to work with
sections of this bundle; instead
the points
along each fiber are identified (much as physically-equivalent
gauge fields are identified in this article).

The projection matrix is
\beq
g^{i}_{j} = \delta_{i j}-
\frac{x^{i} x^{j}}{{\bf x} \cdot {\bf x}}\;, \label{A2}
\eeq
which vanishes on the one-dimensional subspace spanned by ${\bf x}$
and gives unity on the space spanned by ${\bf y}$ and ${\bf z}$. This
tensor plays the role of the identity. The reader who understands the
mismatch of higher and lower indices in (\ref{A1}) should
not have any objection in this regard to (\ref{A2}). The inverse
metric tensor is
\beqa
g^{i j}= {\bf x} \cdot {\bf x}\; \delta_{i j}-
x^{i} x^{j}
\;, \nonumber
\eeqa
and (\ref{AA9}) is satisfied.

Naively, the volume element is 
$d^{3}x {\sqrt {det^{\r}\,g}}$. Such reasoning
implies that the total volume is
\beqa
V=\int d^{3}x {\sqrt {det^{\r}\,g}}=\int 
\frac{d^{3}x}{{\bf x}\cdot{\bf x}}
=\infty\;,  \nonumber
\eeqa
The
problem is that singular
directions are being integrated over, and the
integration over these directions produces
infinity. Now if the function
to be integrated does
not depend on 
${\bf x} \cdot {\bf x}$, then there is 
a straightforward way to avoid doing this 
unwanted integration. The correct
measure is obtained by dividing out the
differential in the direction of the singular vector, ${\bf x}$:
\beqa
dV= \frac{d^{3}x {\sqrt {det^{\r}\,g}}}{{\bf x} \cdot d{\bf x}}
=2\frac{d^{3}x {\sqrt {det^{\r}\,g}}}{d({\bf x} \cdot {\bf x})}\;. \nonumber
\eeqa
Formally, this is made somewhat easier by integrating this unwanted
differential
over a normalized function, so that the measure becomes, e.g.
\beq
dV= 2 d^{3}x {\sqrt {det^{\r}\,g}} 
\;({\sqrt {\frac{\pi}{\a}}}e^{-\a {\bf x} \cdot {\bf x}})
\;, \label{A5}
\eeq
where $\a$ is a constant with units of length squared. Upon 
substituting polar coordinates, (\ref{A5}) is
\beqa
dV=2{\sqrt {\frac{\pi}{\a}}}\; \int_{0}^{\infty} dr\; e^{-\a r^{2}}
\int_{-1}^{1} d\cos \theta \int_{0}^{2\pi} d\phi\;, \nonumber
\eeqa
which is clearly correct when integrating over functions which
do not depend on $r$. Alternatively, if one is considering
ratios of integrals on the sphere, the divergent factor from
the unwanted integration (which is an overall constant)
will cancel (though doing this explicitly requires
a cut-off on the measure).

The affine connection (\ref{AA15}) is
\beq
{\G}^{j}_{k\,l}= \frac{2x^{j}x^{k}x^{l}}{({\bf x} \cdot {\bf x})^{2}}
-\frac{\delta_{j\;l}x^{k}+\delta_{j\;k}x^{l}}{{\bf x} \cdot {\bf x}}
\;.  \label{A7}
\eeq

By constructing the vectors
$e^{K}_{k}$, one can show that 
$e^{J}_{l}e^{K}_{m} g^{lm}=\delta^{JK}$, which 
is a constant. This makes explicit the
fact that the covariant derivative
of the
metric tensor vanishes. This can also
be seen by direct calculation:
\beqa
\nabla_{n}g^{j}_{q}=g^{r}_{q} \nabla_{n}g^{j}_{r}
=g^{k}_{l} g_{qs} \partial_{n} g^{ls}+ g^{ks} g^{r}_{q} 
\partial_{n} g_{sr}=g^{s}_{q} (\partial_{n} g^{p}_{s}) g_{pq} \;. \label{A8}
\eeqa
Substituting (\ref{A7}) directly into (\ref{A8}) gives zero.

%The geodesic equation is obtained by extremizing the
%length integral:
%\beqa
%0=\delta \int ds= 
%\delta \; \int {\sqrt {g_{ij}(x(s))\;dx^{i}(s)\,\,dx^{j}(s)} } \;.  \label{A9}
%\eeqa
%The result is identical to that in the case of nonsingular
%metric, namely the second-order differential equation
%\beqa
%\frac{d^{2} x^{j}(s)}{ds^{2}} + {\G}^{j}_{lm}(x(s)) \;\frac{dx^{l}(s)}{ds} 
%\frac{dx^{m}(s)}{ds} =0\;.  \label{A11}
%\eeqa
%Upon
%substituting (\ref{A7}) 
%into this expression, the 
%second term is found to be zero. Therefore, a geodesic
%is described by
%\beqa
%x^{j}=x_{0}^{j}+x_{1}^{j} s\;, \label{A12}
%\eeqa
%where 
%$x_{0}^{j}$ and $x_{1}^{j}$ are constant coefficients. This
%is not a straight line, but part of
%a great circle (to see this, recall that the point on the 
%unit sphere in $R^{3}$ described by $x^{j}$ is ${\hat {\bf x}}=
%\frac{{\bf x}}{\vert {\bf x} \vert}$). Unfortunately, (\ref{A12})
%only describes half a great circle (allowing for $s<0$). It 
%is difficult, but
%not impossible, to construct closed geodesics with the singular metric 
%formalism. This must be done by taking at gluing at least
%three finite segments from such curves 
%(\ref{A12}) together. Such a gluing is also needed
%to find conjugate
%points along geodesics.

The Laplacian (\ref{A10}), with a little work, can
be shown to agree with
the expression of 
the Laplacian on a sphere in polar coordinates:
\beqa
\Delta=-\frac{1}{\sin \theta} \pr_{\theta} \sin \theta \pr_{\theta} 
-\frac{1}{\sin^{2} \theta} \pr_{\phi}^{2} \;, \nonumber
\eeqa
which is left for the reader to verify.

Geodesics on the sphere are described by straight lines in
$R^{3}$, namely ${\bf x}(t)={\bf a}+{\bf b}t$, where 
$t \in (-\infty, \infty)$. Such a curve is mapped
to a great half-circle by 
${\bf x} \rightarrow {\hat {\bf x}}$. It
is impossible to have a conjugate pair of points on this
curve (these are antipodes on the sphere), since these
are approached as 
$t \rightarrow \pm \infty$. However, by connecting
several such curves together, a complete great circle
can be made, and conjugate points can be reached.

\section*{Acknowlegements}

I thank
E. Akhmedov, J. Ambj{\o}rn, M.~N. 
Chernodub, P.~H. Damgaard, M. Kudinov, Y.
Makeenko, M. Polikarpov, G.~K. Savvidy and especially  J.
Huntley and V.~P. Nair for discussions. Pierre van Baal
pointed out a serious error concerning
Chern-Simons integrals in an earlier
version of this paper. Some of this
work was done while visiting
ITEP, whose staff I am
grateful to for
a very pleasant stay in Moscow. Finally, I thank
the staff of the Niels
Bohr Institute for their 
hospitality and financial support.

\vfill

\end{document}